\newcommand{\dm}[1]{\textcolor{black}{#1}}
\newcommand{\dmt}[1]{\textcolor{black}{#1}}

\documentclass[review]{elsarticle}
\usepackage{amsmath,graphicx}
\usepackage{amsmath,amssymb} 
\usepackage{epsfig}
\usepackage{multirow,color}
\usepackage{llncsdoc}
\usepackage{etoolbox}

\usepackage{framed,multirow}
\usepackage{lineno,hyperref}
\usepackage{latexsym}
\usepackage{url}
\usepackage{algorithm}
\usepackage[noend]{algpseudocode}
\makeatletter
\def\BState{\State\hskip-\ALG@thistlm}
\makeatother

\begin{document}

\begin{frontmatter}


\title{Generative Adversarial Networks And Domain Adaptation For Training Data Independent Image Registration}



\author{Dwarikanath Mahapatra}
\address{Inception Institute of Artificial Intelligence, Abu Dhabi, UAE \\
\textit{dwarikanath.mahapatra@inceptioniai.org}
}


\begin{abstract}
Medical image registration is an important task in automated analysis of \dm{multimodal images} and temporal data involving multiple patient visits. Conventional approaches, although useful for different image types, are time consuming. Of late, deep learning (DL) based image registration methods have been proposed that outperform traditional methods in terms of accuracy and time. However, DL based methods are heavily dependent on training data and do not generalize well when presented with images of different scanners or anatomies. We present a DL based approach that can perform medical image registration of one image type despite being trained with images of a different type. This is achieved by unsupervised domain adaptation in the registration process and allows for easier  application to different datasets without extensive retraining. 
To achieve our objective we train a network that transforms the given input image pair to a latent feature space vector using autoencoders. The resultant  encoded feature space is used to generate the registered images with the help of generative adversarial networks (GANs). This feature transformation ensures greater invariance to the input image type. Experiments on chest Xray, retinal and brain MR images show that our method, \dm{trained on one dataset} gives better registration performance \dm{for other datasets, outperforming } conventional methods that do not incorporate domain adaptation.
\end{abstract}

 
\end{frontmatter}


\section{Introduction}
\label{sec:intro}

Elastic (or deformable) image registration is crucial for \dmt{many} medical image analysis tasks such as computer aided diagnosis, atlas building, correcting deformations and monitoring pathological changes over time. It is especially relevant in monitoring patients over multiple visits and quantifying change of tumor volume and other pathologies. Such quantification gives an accurate estimate of treatment effectiveness. Deformable registration establishes a non-linear correspondence between a pair of images or 3-D image volumes. Conventional registration methods solve an optimization problem for each volume pair that aligns voxels with similar appearance while enforcing smoothness constraints on the registration mapping. This requires iterative gradient descent methods which are computationally intensive, and extremely slow in practice.

\dmt{With the increasing number of works focusing on deep learning (DL) based image analysis, many interesting works on image registration have also been proposed. The method by} \cite{Mahapatra_CVIU2019,Mahapatra_PR2020,Mahapatra_CMIG2019,Mahapatra_LME_PR2017,Zilly_CMIG_2016,BalaCVPR18} learns a parameterized registration function from a collection of volumes, \dm{while there are other methods that use regression or generative models \cite{RegNet,Mahapatra_MLMI2018,ZGe_MTA2019,Mahapatra_SSLAL_CD_CMPB,Mahapatra_SSLAL_Pro_JMI,Mahapatra_LME_CVIU,LiTMI_2015}}. The primary shortcoming of current DL based image registration methods is a reliance on large \dm{ \emph{labeled}} datasets for training. Although conventional registration methods are time consuming, they perform equally well on different types of image pairs. However, a DL network trained to register a pair of brain magnetic resonance (MR) images does not  perform convincingly on a pair of chest xray images or vice-versa. Similar outcomes are observed if the test image is of the same anatomy but acquired using a different scanner. As a result, DL based image registration methods have to be retrained when presented with a new dataset. In general, it is challenging to obtain ground truth data for registration which restricts the method's use on new image types and in real world scenarios. In this paper we address the problem of using \dmt{pre-trained} registration networks on new datasets by proposing a domain adaptation based  DL method that, once trained on a particular dataset, can be easily used on other image pairs with minimal retraining.

The dataset independent registration is achieved by \dmt{first} using a network trained on a different dataset having ground truth correspondences for registration between source and target images. To enable transfer of knowledge across domains we use a latent feature space representation of the images and use them to generate the registered image instead of the original images. The latent feature representation is obtained from autoencoders and then fed to a generator network that generates the registered image and the deformation field. \dmt{Previous work in \cite{PR_Falez,MahapatraJDI_Cardiac_FSL,Mahapatra_JSTSP2014,Behzad_PR2020,MahapatraTIP_RF2014,MahapatraTBME_Pro2014} demonstrates that between autoencoders and spiking neural networks, autoencoders do a better job for unsupervised visual feature learning. Hence we use them for obtaining a latent space representation with sufficiently discriminative features.} The generated images are used as input to a discriminator that determines its accuracy and realism. 

\subsection{Related Work}

\dmt{We refer the reader to \cite{RegRev,MahapatraTMI_CD2013,MahapatraJDICD2013,MahapatraJDIMutCont2013,MahapatraJDIGCSP2013,MahapatraJDIJSGR2013} for a comprehensive review on conventional image registration methods. Although widely used, these methods are very time consuming, especially when registering image volumes.} This is due to: 1) the iterative optimization techniques common to all methods; and 2) extensive parameter tuning involved. On the other hand, DL methods \dmt{can potentially overcome}  these limitations by using trained models to output the registered images and the deformation field in much less time.

 Since we propose a deep learning approach we review image registration methods that use it as part of their pipeline. 
Wu et. al. \cite{WuTBME,MahapatraJDISkull2012,MahapatraTIP2012,MahapatraTBME2011,MahapatraEURASIP2010}  used convolutional stacked autoencoders (CAE) to extract features from fixed and moving images, which are used in a conventional iterative deformable registration framework.
Miao et. al. \cite{Miao_Reg,Kuanar_ICIP19,Bozorgtabar_ICCV19,Xing_MICCAI19,Mahapatra_ISBI19,MahapatraAL_MICCAI18} use convolutional neural network (CNN) regressors to rigidly register synthetic images. Liao et. al. \cite{Liao_Reg,Mahapatra_MLMI18,Sedai_OMIA18,Sedai_MLMI18,MahapatraGAN_ISBI18,Sedai_MICCAI17} use CNNs and reinforcement learning for iterative registration of CT to cone-beam CT in cardiac and abdominal images. FlowNet \cite{FlowNet,Mahapatra_MICCAI17,Roy_ISBI17,Roy_DICTA16,Tennakoon_OMIA16,Sedai_OMIA16} formulates dense optical flow estimation as a regression task using CNNs and uses it for image matching.
%
%
%
These approaches still use a conventional model to generate the transformed image from the deformation field which increases computation time and does not fully utilize the generative capabilities of DL methods \dmt{for the purpose of generating} registered images.

Jaderberg et al. \cite{STN,Mahapatra_MLMI16,Sedai_EMBC16,Mahapatra_EMBC16,Mahapatra_MLMI15_Optic,Mahapatra_MLMI15_Prostate} introduced spatial transformer networks (STN) to align input images in a larger task-specific network. STNs, however, need many labeled training examples \dmt{which makes them unsuitable for our application}. 
%
%
Sokooti et. al. \cite{RegNet,Mahapatra_OMIA15,MahapatraISBI15_Optic,MahapatraISBI15_JSGR,MahapatraISBI15_CD,KuangAMM14} propose RegNet that uses CNNs trained on simulated deformations to generate displacement vector fields for a pair of unimodal images.
 Vos et. al. \cite{Vos_DIR,Mahapatra_ABD2014,Schuffler_ABD2014,MahapatraISBI_CD2014,MahapatraMICCAI_CD2013,Schuffler_ABD2013} propose the deformable image registration network (DIRNet) which takes pairs of fixed and moving images as input, and outputs a transformed image non-iteratively. Training is completely unsupervised and unlike previous methods it is not trained with known registration transformations. 
 Uzunova et. al. \cite{UzunovaMICCAI2017,MahapatraProISBI13,MahapatraRVISBI13,MahapatraWssISBI13,MahapatraCDFssISBI13,MahapatraCDSPIE13} propose to learn representative shape and appearance models from few training samples, and embed them in a new model-based data augmentation scheme to generate huge amounts of ground truth data. 
Rohe et. al. \cite{RoheMICCAI2017,MahapatraABD12,MahapatraMLMI12,MahapatraSTACOM12,VosEMBC,MahapatraGRSPIE12} propose SVF-Net that trains a network using reference deformations obtained by registering \emph{previously segmented} regions of interest (ROIs). 
Balakrishnan et. al. \cite{BalaCVPR18,MahapatraMiccaiIAHBD11,MahapatraMiccai11,MahapatraMiccai10,MahapatraICIP10,MahapatraICDIP10a} proposed a CNN based method that learns a parameterized registration function from a collection of training volumes, and does not require ground truth correspondences during training. The above methods are limited by the need of spatially corresponding patches (\cite{RegNet,Vos_DIR,MahapatraICDIP10b,MahapatraMiccai08,MahapatraISBI08,MahapatraICME08,MahapatraICBME08_Retrieve,MahapatraICBME08_Sal,MahapatraSPIE08,MahapatraICIT06}) or being too dependent on training data.

The above limitations can be overcome to some extent by using generative models that directly generate the registered image and the deformation field. In previous work \cite{MahapatraGANISBI2018,sZoom_Ar,CVIU_Ar,AMD_OCT,GANReg1_Ar,PGAN_Ar} we used generative adversarial networks (GANs) for multimodal retinal image registration where the trained network outputs the registered image and the deformation field. \dm{We show that by incorporating appropriate constraints in the adversarial loss and content loss function, generative models are fairly reliable in directly generating the registered image and its corresponding deformation field. However, we also observe that GANs tend to misregister local regions with multiple structures. To address this shortcoming,} in \cite{Mahapatra_MLMI2018,Haze_Ar,Xr_Ar,RegGan_Ar,ISR_Ar,LME_Ar} we propose a joint registration and segmentation method that incorporates segmentation information into the registration process to achieve better registration performance than conventional methods. \dm{The segmentation information is integrated as part of the adversarial loss function and experimental results show its significant contribution in improving registration accuracy}.

Despite the improvements brought about by joint registration and segmentation in \cite{Mahapatra_MLMI2018,Misc,Health_p,Pat2,Pat3,Pat4}, its success depends upon training data with corresponding information about the applied deformation field. Consequently, this restricts it's application due to the need for generating images with known deformation field. 
\dm{Previous works have explored different feature spaces for domain independent tasks as in cross modality face detection  using deep local descriptors and cross modality enumeration loss \cite{PR_DLFace,Pat5,Pat6,Pat7,Pat8,Pat9}. This approach has led to better robustness on traditional datasets. Shang et. al. in \cite{PR_Shang,Pat10,Pat11} propose a unsupervised feature selection algorithm for sparse subspace learning that simultaneously preserves local discriminant structure and local geometric structure. Li et. al. \cite{LiTMI_2015} propose auto correlation based local landmark alignment. 
Yang et. al. in \cite{PR_Yang} propose an integrated tumor classification framework that  addresses problems of small sample and unbalanced datasets using an inverse space sparse representation.
}
\dm{
With popularity of deep learning methods much focus has also been put on efficient learning approaches such as class specific features \cite{PR_Lei}, asymmetric 3D convolutions \cite{PR_HYang}, and semi-supervised domain adaptation \cite{PR_WWang}, while autoencoders have been used in \cite{PR_LHou} for simultaneous nucleus detection and feature extraction in histopathology tissue images.
}
To overcome the challenge of transforming medical images to a suitable feature space, we explore a domain adaptation based method that uses a  pre-trained network on one type of images to register novel images of another image type.

Compared to our previous works (\cite{MahapatraGANISBI2018,Mahapatra_MLMI2018}) this work has the following novelties: 1) we use latent vector representation from convolutional auto encoders in the image registration process; 2) the latent space image is used to generate registered image and the deformation field; 3) the latent feature representation also serves the purpose of domain adaptation wherein a new image pair is easily registered with no network finetuning.

%

\section{Methods}
\label{sec:met}

\begin{figure}[t]
\begin{tabular}{c}
\includegraphics[height=3.9cm, width=7.99cm]{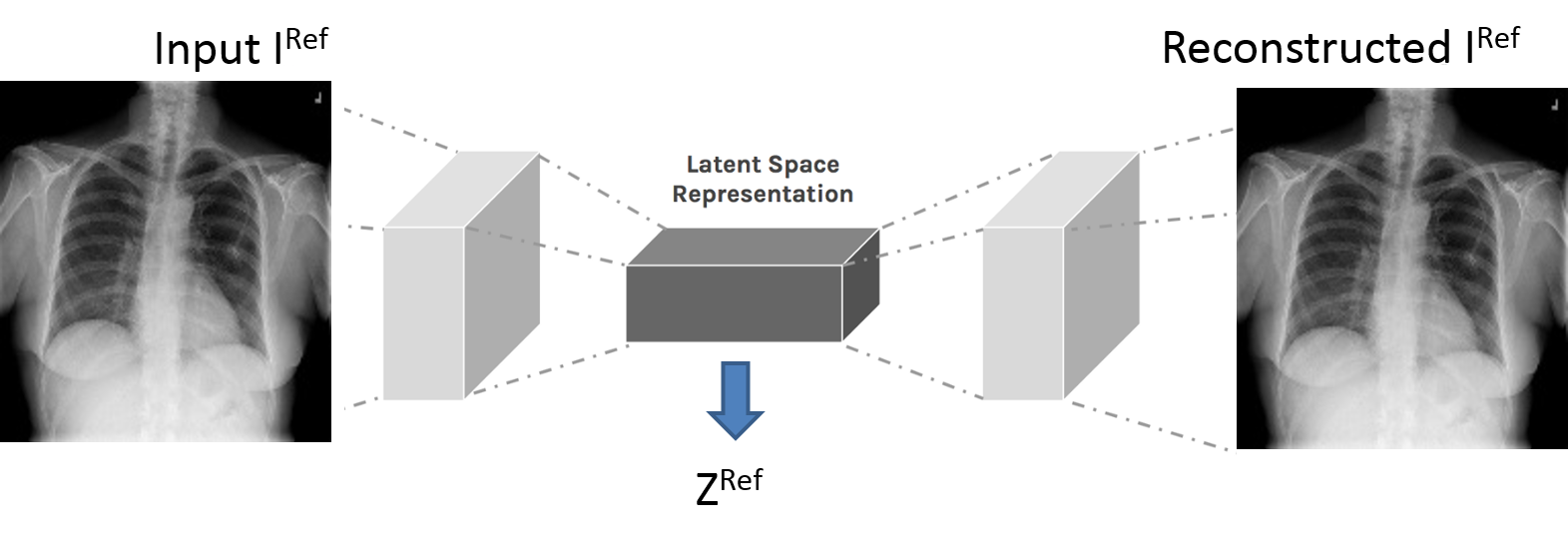}  \\ 
(a) \\
\includegraphics[height=3.9cm, width=7.99cm]{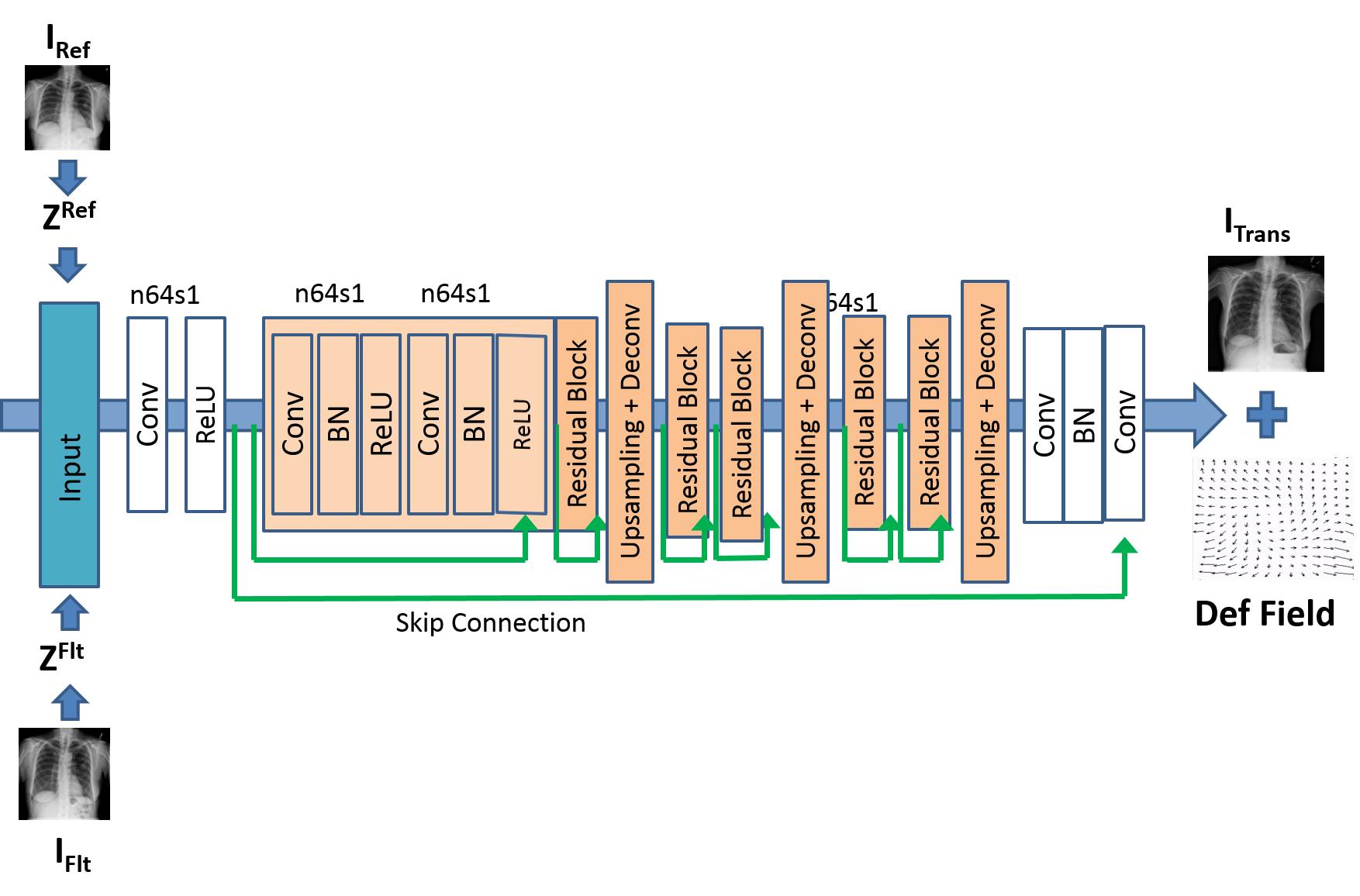}  \\ 
(b) \\
\includegraphics[height=3.9cm, width=7.99cm]{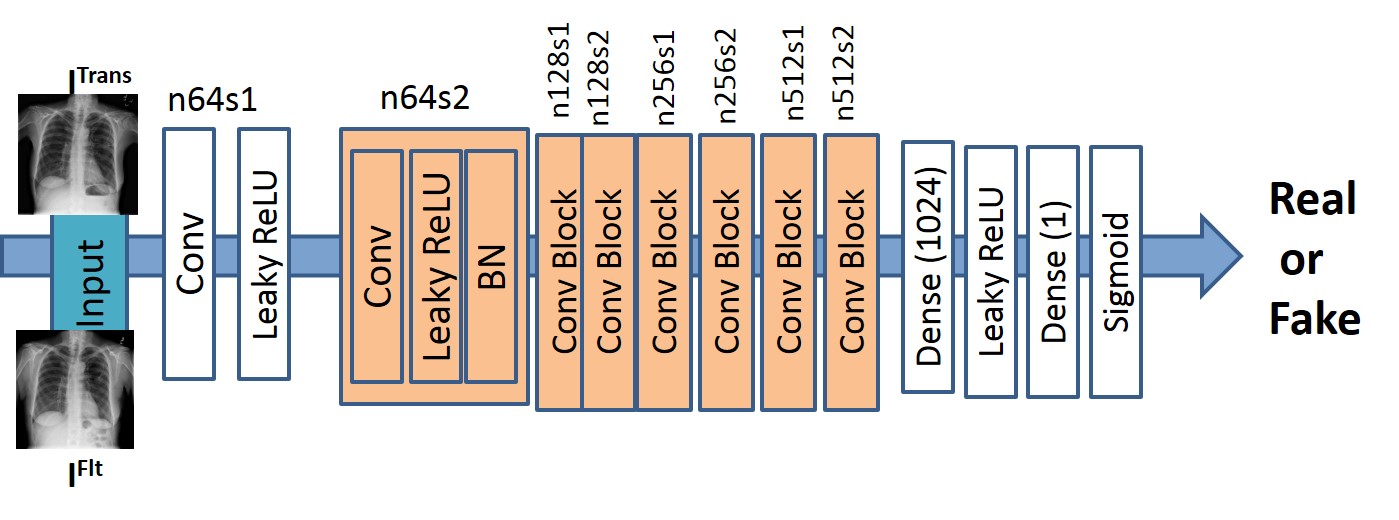}  \\
(c)\\
\end{tabular}
\caption{(a) Convolutional Autoencoder architecture; (b) Generator Network; (c) Discriminator network. $n64s1$ denotes $64$ feature maps (n) and stride (s) $1$ for each convolutional layer.}
\label{fig:diagram1}
\end{figure}

Figure~\ref{fig:diagram1} shows the architecture of our proposed method. In the first stage  (Figure~\ref{fig:diagram1} (a) ) two convolutional autoencoders (CAEs) are trained on the reference image ($I^{Ref}$) and floating image ($I^{Flt}$). The CAEs are basically tasked with reconstructing the input image through a series of encoding and decoding stages. The output of the encoding stage results in a \dmt{2D} latent space representation (which we denote as $z^{src}$ and $z^{flt}$ respectively) that is used to reconstruct the original image. The latent space representations act as input to a generative adversarial network (GAN) whose generator network reconstructs the original image (Figure~\ref{fig:diagram1} (b) ) through a series of convolution and upsampling  stages. The output of the generator network  is the registered image or transformed image ($I^{Trans}$) and the recovered deformation field ($I^{Def-Recv}$). \dm{The output matrix shape has same number of rows as input image, but twice the number of columns, placing the registered image and deformation field side by side. The first half-image is the registered image while the second half-image is the deformation field.}  During training the discriminator network (Figure~\ref{fig:diagram1} (c) ) also comes into play by taking as input $I^{Reg}$ and $I^{Def-Recv}$  and comparing with the ground truth data. The feedback from the discriminator influences the updates of the generator weights. During testing only the generator network is used. \dmt{The steps for training the registration network is summarized in Algorithm~$1$, and subsequent steps to obtain a registered image during inference time is summarized in Algorithm~\ref{algtest}.}

GANs \cite{GANs} are generative models trained in an adversarial setting. The generator $G$ outputs a desired image type while discriminator $D$ outputs a probability of the generated image matching the training data. GANs have been used in various applications such as image super resolution \cite{SRGAN,MahapatraMICCAI_ISR,Mahapatra_CMIG2019}, image synthesis and image translation using conditional GANs (cGANs) \cite{CondGANs,Mahapatra_CVIU2019,MahapatraAL_MICCAI18} and cyclic GANs (cycleGANs) \cite{CyclicGANs}. 
In cGANs the output is conditioned on the input image and a random noise vector, and requires training image pairs. On the other hand cycleGANs do not require training image pairs but enforce consistency of deformation field. 
%
%

%
%

\begin{algorithm}{}
\begin{algorithmic}[1]
\Procedure{Domain Adaptation Based Image Registration}{}
\State $I^{Ref}$,$I^{Flt}$ $\gets$ \textit{Reference and Floating Images}
\State $z^{Ref}$,$z^{Flt}$ $\gets$-convolutional auto encoder outputs for $I^{Ref}$,$I^{Flt}$ 
\State $I^{Def-App}$  $\gets$ actual deformation field between $I^{Ref}$ and $I^{Flt}$
\BState \emph{Training loop}:
\State Input images  $I^{Ref}$,$I^{Flt}$ are fed into the Generator $G$
\State Auxillary inputs $z^{Ref}$,$z^{Flt}$ to $G$ 
\State Generate intermediate registered image  $I^{Trans}$ and recovered deformation field $I^{Def-Recv}$
\State Feed $I^{Trans}$,$I^{Ref}$,$I^{Flt}$,$I^{Def-App}$,$I^{Def-Recv}$ to discriminator $D$ 
\State Calculate $l_{content}$ (Eqn.~\ref{eq:conLoss}) and $L_{adv}$ (Eqn.~\ref{eqn:adloss}), and combine to get the full objective function term (Eqn.~\ref{eqn:cyGan3})
\State \textbf{close}
\State \textbf{Outputs}: Final $UReg-Net$ to output $I^{Trans}$ and  $I^{Def-Recv}$
\EndProcedure
\caption{Training Steps for Registration Network}
\end{algorithmic}
\label{algtr}
\end{algorithm}

\begin{algorithm}{}
\label{algtest}
\begin{algorithmic}[1]
\Procedure{Domain Adaptation Based Image Registration}{}
\State $I^{Ref}$,$I^{Flt}$ $\gets$ \textit{Reference and Floating Images} for registration
\State Obtain $z^{Ref}$,$z^{Flt}$ $\gets$ CAE outputs for $I^{Ref}$,$I^{Flt}$ 
\State Input  $z^{Ref}$,$z^{Flt}$ to $UReg-Net$
\State Output $I^{Trans}$ and $I^{Def-Recv}$, the registered image and recovered deformation field 
\State Calculate registration error between $I^{Def-Recv}$ and $I^{Def-App}$ (if available)
\EndProcedure
\caption{Register Image Pair Using Pre-trained $UReg-Net$}
\end{algorithmic}
\end{algorithm}

\subsection{Convolutional Autoencoder Training}

Our proposed convolutional auto encoder (CAE) (Figure~\ref{fig:diagram1} (a)) takes a $512\times512$  input image. The first convolution layer has $32$ filters of size $3\times3$ followed by a max pooling layer of subsampling factor $2$. This is followed by two more stages of convolution and max pooling layers. The two convolution layers have $64$ filters each of $3\times3$. Thus the encoder part outputs a $2D$ latent space representation of the input image denoted as $z$ \dm{ of dimension $64\times64$}. Similarly the decoder has 3 stages of upsampling and deconvolution  layers that reconstruct the original image. 
 %
%
 In the first stage we train the CAE end to end to reconstruct the original image and consequently $z$ is an accurate representation of the input image in the latent space. We train two separate CAEs for $I^{Ref}$ and $I^{Flt}$. Since we use all convolution layers our network is independent of input image size, and the number of layers in the encoder and decoder part can be varied to suit our training data.

\subsection{Generating Registered Images}

Figure~\ref{fig:diagram1} (b) shows the generator network $G$. It employs residual blocks, each block having two convolution layers with $3\times3$ filters and $64$ feature maps, followed by batch normalization and ReLU activation.  Each convolution layer is followed by a upsampling layer and deconvolution. There are three stages of convolution and upsampling blocks thus restoring the latent space representation back to its original dimension.  In addition to generating the registered image $I^{Trans}$, $G$ also outputs a deformation field $I^{Def-Recv}$.

The training database has chest Xray images of patients with multiple visits. To generate training data each of the images is subjected to various types of local elastic deformation using B-splines with the pixel displacements in the range of $\pm[1,20]$.
The original undistorted image is $I^{Ref}$ and the resulting deformed images are $I^{Flt}$. Each $I^{Ref}$ can have multiple $I^{Flt}$.  Applying synthetic deformations allows us to: 1) accurately quantify the registration error between $I^{Def-App}$ and $I^{Def-Recv}$, the applied and recovered deformation fields; and 2) determine the similarity between $I^{Trans}$ and $I^{Ref}$.

 

%
%
$G$ is a feed-forward CNN whose parameters $\theta_G$ are given by, 
\begin{equation}
\widehat{\theta}=\arg \min_{\theta_G} \frac{1}{N} \sum_{n=1}^{N} l^{Reg}\left(G_{\theta_G}(I^{Flt}),I^{Ref},I^{Flt}\right),
\label{eq:theta1}
\end{equation}
where the registration loss function $l^{Reg}$ combines content loss (Eqn.~\ref{eq:conLoss}) and adversarial loss (Eqn.~\ref{eqn:cyGan1}), and $G_{\theta_G}(I^{Flt})=I^{Trans}$ the registered image. The content loss is given by 
\begin{equation}
\begin{split}
& l_{content}(I^{Trans},I^{Ref}) = NMI (I^{Ref},I^{Trans}) \\
& + \left[1-SSIM(I^{Ref},I^{Trans})\right] + VGG(I^{Ref},I^{Trans}). 
\end{split}
\label{eq:conLoss}
\end{equation} 

$NMI$ denotes normalized mutual information between $I^{Ref}$ and $I^{Trans}$ and  is suitable for multimodal and unimodal deformable registration \cite{FFD}. \dm{NMI determines the joint occurence of intensity pairs for corresponding voxels across an image pair.  NMI is robust in accounting for intensity changes and is robust to any intensity distribution differences that may arise due to intensity variations as a result of view point variance.} 
 $SSIM$ denotes structural similarity index metric (SSIM) based on edge distribution \cite{SSIM} and quantifies landmark correspondence between different images. $SSIM\in[0,1]$ with higher values indicating greater similarity. \dm{The advantage of SSIM is independence from intensity matching, as it relies on shape matching which is more robust under different conditions. } 

$VGG$ is the $L2$ distance between two images using all the feature maps obtained from a pre-trained $VGG16$ network \cite{VGG}. This sums up to $64\times2+128\times2+256\times2+512\times3+512\times3=3968$ feature maps. The feature maps are of different dimensions due to multiple max pooling steps. Using all feature maps ensures we are comparing information from multiple scales, both coarse and fine, and thus improves robustness. The feature maps are normalized to have values between $[0,1]$. 

The discriminator $D$ (Figure~\ref{fig:diagram1} (c)) has eight convolution layers with the kernels increasing by a factor of $2$ from $64$ to $512$ . Leaky ReLU activation is used and strided convolutions reduce the image dimension  when the number of features is doubled. The resulting $512$ feature maps are
followed by two dense layers and a final sigmoid activation to obtain a probability map. $D$ evaluates similarity of intensity distribution between $I^{Trans}$ and $I^{Flt}$, and the error between generated and reference deformation fields. 




\subsection{Deformation Field Consistency}

CycleGANs \cite{CyclicGANs} learn mapping functions $G : X \rightarrow Y$ and $F : Y \rightarrow X$,  between image sets $X=I^{Flt}$ and $Y=I^{Ref}$. Adversarial discriminators $D_X$ differentiate  between images $x$ and registered images ${F(y)}$, and $D_Y$ distinguishes between ${y}$ and ${G(x)}$. 
$G$ registers $I^{Flt}$ to $I^{Ref}$ while $F$ is trained to register $I^{Ref}$ to $I^{Flt}$. We refer the reader to \cite{CyclicGANs} for implementation details.
In addition to the content loss (Eqn~\ref{eq:conLoss}) we have: 1) an adversarial loss; 
and 2) a cycle consistency loss to ensure transformations $G,F$ do not contradict each other.

\paragraph{\textbf{Adversarial Loss}:}

The adversarial loss is an important component to ensure that the generated outputs are plausible. In previous applications the adversarial loss was based on the similarity of generated image to training data distribution. Since our generator network has multiple outputs we have additional terms for the adversarial loss. The first term matches the distribution of $I^{Trans}$ to  $I^{Flt}$  
and is given by:
%
\begin{equation}
L_{cycGAN}(G,D_Y) = E_{y\in p_{data}(y)} \left[\log D_Y(y)\right] + E_{x\in p_{data}(x)} \left[\log \left(1-D_Y(G(x))\right)\right], 
\label{eqn:cyGan1}
\end{equation}
We retain notations $X,Y$ for conciseness.
%
 There also exists $L_{cycGAN}(F,D_X)$, the corresponding adversarial loss for $F$ and $D_X$ and is given by  
%
\begin{equation}
L_{cycGAN}(F,D_X) = E_{x\in p_{data}(x)} \left[\log D_X(x)\right] + E_{y\in p_{data}(y)} \left[\log \left(1-D_X(F(y))\right)\right].
\label{eqn:cyGan12}
\end{equation}

%
Cycle consistency loss ensures that for each $x \in X$ the reverse deformation should bring $x$ back to the original image, i.e. $x \rightarrow G(x) \rightarrow F(G(x))\approx x$. 
Similar constraints also apply for mapping $F$ and $y$. This is achieved using, 
%
%
\begin{equation}
L_{cyc}(G,F)= E_{x} \left\|F(G(x))-x\right\|_1 + E_{y} \left\|G(F(y))-y\right\|_1,
\label{eqn:cyGan2}
\end{equation}
%
%
%
\dm{
Instead of the original images, $I^{Ref}$ and $I^{Flt}$, we use their latent vector representations as input to the generator to obtain $I^{Trans}$. 
}

\subsection {Integrating Deformation Field Information}

The second component of the adversarial loss is 
 the mean square error between $I^{Def-App}$ and $I^{Def-Recv}$, the applied and recovered deformation fields. 
The final adversarial loss is 
\begin{equation}
\begin{split}
L_{adv}= & L_{cycGAN}(G,D_{I^{Ref}}) + L_{cycGAN}(F,D_{I^{Flt}}) \\  
& + \log \left(1-MSE_{Norm}(I^{Def-App},I^{Def-Recv}) \right), 
\end{split}
\label{eqn:adloss}
\end{equation}
where $MSE_{Norm}$ is the MSE normalized to $[0, 1]$, and $1-MSE_{Norm}$ ensures that similar deformation fields gives a corresponding higher value. Thus all terms in the adversarial loss function have values in $[0,1]$. 
 The full objective function is given by 
\begin{equation}
L(G,F,D_{I^{Flt}},D_{I^{Ref}})= L_{adv} + l_{content} + \lambda L_{cyc}(G,F) 
\label{eqn:cyGan3}
\end{equation}
where $\lambda=10$ controls the contribution of the two objectives. The optimal parameters are given by:
\begin{equation}
G^{*},F^{*}=\arg \min_{F,G} \max_{D_{I^{Flt}},D_{I^{Ref}}} L(G,F,D_{I^{Flt}},D_{I^{Ref}})
\label{eqn:CyGan4}
\end{equation}

\subsection{Registering A Test Image}
\label{met:newimg}

We have trained our image with reference and floating images from the same modality (i.e., chest xray). After training is complete the network weights are such that the generated image is the registered version of $I^{Flt}$. A notable difference from previous methods \cite{Mahapatra_MLMI2018,MahapatraGANISBI2018} is instead of the actual images, the input is a latent space representation of the input image pair. This novelty allows us to register different image types by inputting their embedded representations instead of the actual image. Since the network is not trained on the actual images it is agnostic to the image type and can be used for different image modalities. 

In this work we consider two different test image scenarios: 1) brain MR image registration where the reference and floating images are of the same modality; 2) multimodal retinal image registration where the reference image consists of retinal fundus images and the floating images are fluorescein angiography (FA) images. 
For brain MRI registration $I^{Ref}$ is each of the brain image slices while  $I^{Flt}$ is generated through synthetic deformations.  $I^{Ref}$ and $I^{Flt}$ are input through the respective CAEs to obtain $z^{Ref}$ and $z^{Flt}$. Using the latent space representations the network $G$ is able to generate the $I^{Trans}$ and the deformation field $I^{Def-Recv}$. Since we use all convolution layers our network is independent on the input size. 

In the case of multimodal retinal image registration $I^{Ref}$ are the fundus images while the FA images are $I^{Flt}$. They are put through the corresponding pre-trained CAEs and the output $z$ are input to $G$ to obtain the registered image and deformation field. During the test stage the discriminator network $D$ is not used.  
The successful registration of multimodal images and organs different from the training data shows that our method is able to achieve domain adaptation in an unsupervised manner.


\section{Experiments}
\label{sec:expts}

\subsection{Dataset Description}

Our registration method was first validated on the NIH ChestXray14 dataset \cite{NIHXray}. The original dataset contains $112,120$ frontal-view X-rays with $14$
disease labels (multi-labels for each image), which are obtained from associated radiology reports. Since the original dataset is designed for classification studies, we selected samples and applied the following steps to make it suitable for validating registration experiments. 
\begin{enumerate}
\item $50$ patients each from all the $14$ disease classes were selected, giving a total of $14\times50=700$ different patients. Care was taken to ensure that all the patients had multiple visits (minimum $3$ visits and maximum $8$ visits). 


\item As decribed previously we take these images as $I^{Ref}$ and generate corresponding floating images by simulating elastic deformations. 

\item In total we had $2812$ reference images from $700$ patients. For each $I^{Ref}$ we generate $20$ $I^{Flt}$ to obtain $56240$ floating images. 


\item For all our experiments we split the dataset into training, validation and test sets comprising of $70,10,20 \%$ of the images. The split was done at the patient level such that images from a single patient were in one fold only. All the reported results for chest xray images are for the test set. 

\end{enumerate}

Validation of registration performance is usually based on the accuracy of landmark alignment before and after registration. Since we had available the applied deformation field we calculate the alignment/overlap error before registration. Successful registration reduces this error. The registration error is quantified by the mean absolute distance (MAD) which measures the average distance between the two set of edge points for an anatomical region. We manually outline the two lungs in some images using which a segmentation overlap measure such as using Dice Metric (DM) is calculated. Alignment accuracy is also determined using the $95\%$ Hausdorff Distance ($HD_{95}$). An essential  objective of registration is to monitor changes of pathology and organ function over time \dmt{and $HD_{95}$ and DM quantify the overlap of patient segmentation masks from different visits}. 
Lower values of MAD and $HD_{95}$, and higher values of DM indicate better registration performance.
%

\subsection{Implementation Steps}
Our method was implemented in TensorFlow. We use Adam \cite{Adam} with $\beta_1=0.93$ and batch normalization. The generator network $G$ was trained with a learning rate of $0.001$ and $10^{5}$ update iterations. Mean square error (MSE) based ResNet was used to initialize $G$. The final GAN was trained with $10^{5}$ update iterations at learning rate $10^{-3}$. Training and test was performed on a NVIDIA Tesla K$40$ GPU with $12$ GB RAM. 

We show results for: 1) $UReg-Net$: - our proposed unsupervised  registration network; 2) $UReg_{Image}$: - $UReg-Net$ without using the latent space feature $z$ but the original image; 3) $FlowNet$: - the registration method of \cite{FlowNet}; 4) $DIRNet$: - the method of \cite{Vos_DIR}; 5) $JRS-Net$: - the joint registration and segmentation network using GANs (\cite{Mahapatra_MLMI2018}); 6) $Voxel~Morph$: - the registration method of \cite{BalaCVPR18};  and 7) a conventional registration method Elastix \cite{Elastix}. 

Note that only localized deformations were simulated and hence we do not perform any affine alignment before using the different methods. The average training time for the augmented dataset with $56240$ images is $30$ hours. 
The following parameter settings were used for Elastix: non rigid registration using normalized mutual information (NMI) as the cost function. Nonrigid transformations are modelled by B-splines \cite{FFD}, embedded in a multi-grid setting. The grid spacing was set to $ 80,40,20,10,5$ mm with the corresponding downsampling factors being $4, 3, 2, 1, 1$.

\subsection{Chest XRay Image Registration Results}

%
In this section we show results when the trained network is applied on a separate test set of chest xray images consisting of $I^{Ref}$ and $I^{Flt}$ obtained by synthetic deformations.
Registration results of the lung for normal and diseased cases are summarized in Tables~\ref{tab:NIH1},\ref{tab:NIH12}. We observe that the performance is similar for the two kinds of images thus demonstrating that our method is equally effective for diseased and normal images. 
%

\begin{figure}[t]
\begin{tabular}{ccc}
\includegraphics[height=2.5cm,width=2.5cm]{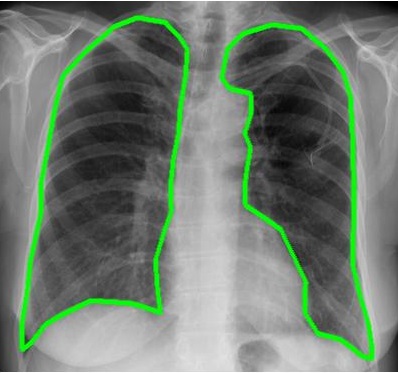} &
\includegraphics[height=2.5cm,width=2.5cm]{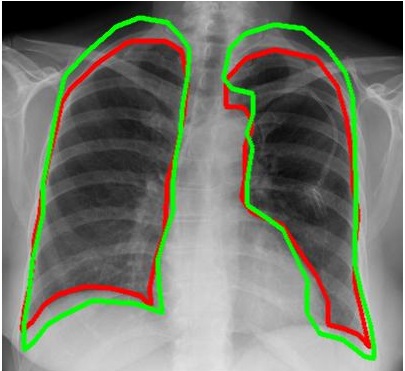} &
\includegraphics[height=2.5cm,width=2.5cm]{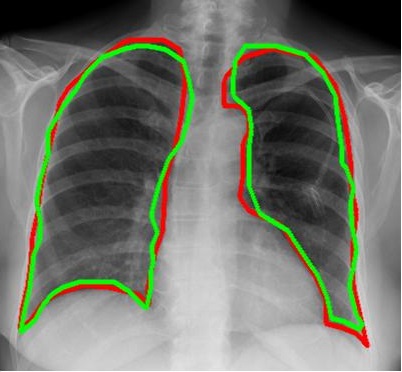} \\
(a) & (b) & (c)  \\
\includegraphics[height=2.5cm,width=2.5cm]{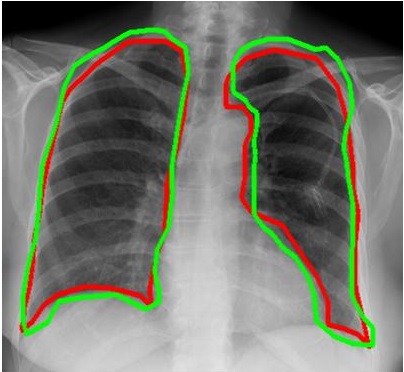} &
\includegraphics[height=2.5cm,width=2.5cm]{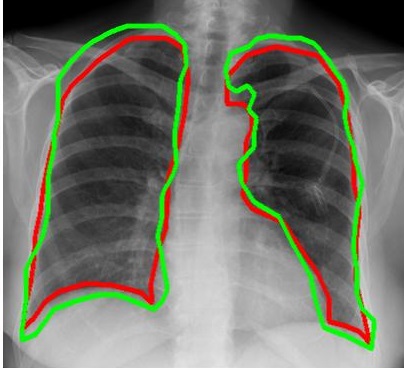} &
\includegraphics[height=2.5cm,width=2.5cm]{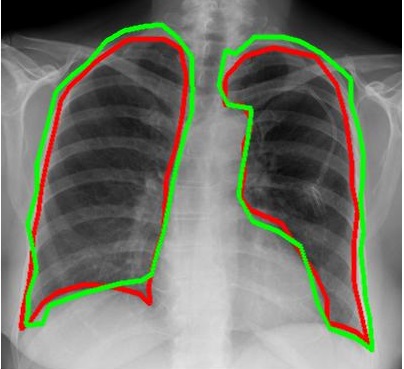}  \\
(d) & (e) & (f) \\ 
\end{tabular}
\caption{Results for \textbf{non-diseased} lung Xray image registration from NIH dataset (patient $5$). (a) $I_{Flt}$ with $I_{Flt}^{Seg}$ (green); (b) $I_{Ref}$ with $I_{Ref}^{Seg}$ (red) and $I_{Flt}^{Seg}$ before registration; Superimposed registered mask obtained using: (c) $UReg-Net$; (d) $DIR-Net$ ; (e) $JRS-Net$; (f) Elastix.}
\label{fig:seg1}
\end{figure}

\begin{figure}[t]
\begin{tabular}{ccc}
\includegraphics[height=2.5cm,width=2.5cm]{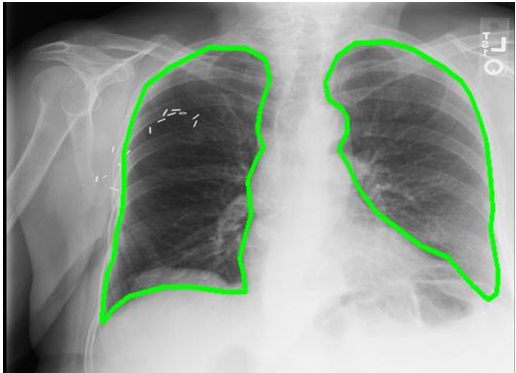} &
\includegraphics[height=2.5cm,width=2.5cm]{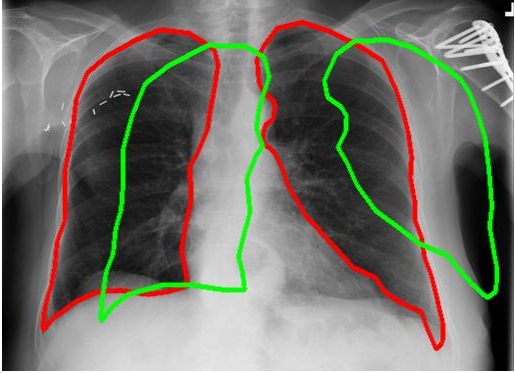} &
\includegraphics[height=2.5cm,width=2.5cm]{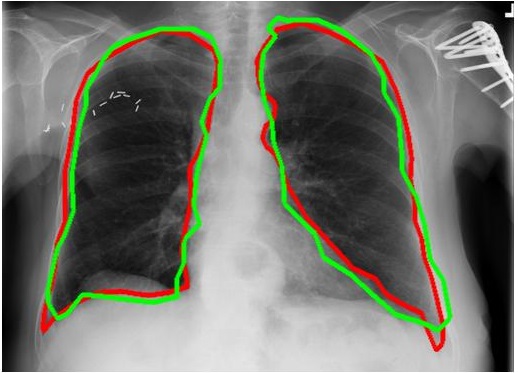} \\
(a) & (b) & (c)  \\
\includegraphics[height=2.5cm,width=2.5cm]{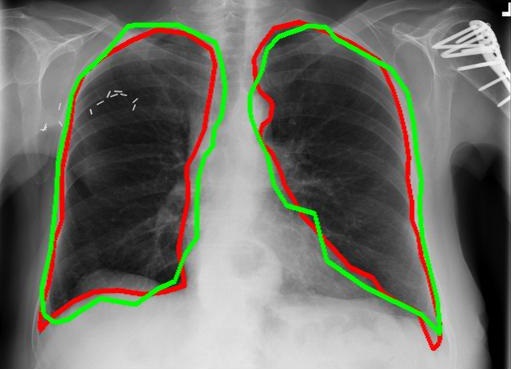} &
\includegraphics[height=2.5cm,width=2.5cm]{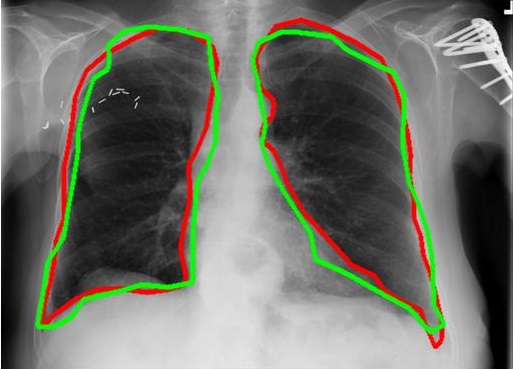} &
\includegraphics[height=2.5cm,width=2.5cm]{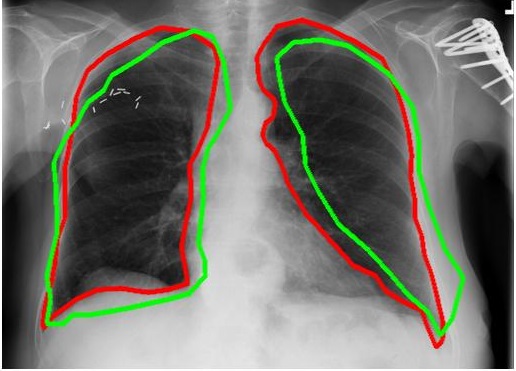} \\
(d) & (e) & (f) \\ 
\end{tabular}
\caption{Results for \textbf{diseased} lung Xray images from NIH dataset (patient $5$). (a) $I_{Flt}$ with $I_{Flt}^{Seg}$ (green); (b) $I_{Ref}$ with $I_{Ref}^{Seg}$ (red) and $I_{Flt}^{Seg}$ before registration; Superimposed registered mask obtained using: (c) $UReg-Net$; (d) $DIR-Net$ ; (e) $JRS-Net$; (f) $Elastix$.}
\label{fig:seg2}
\end{figure}

Figure \ref{fig:seg1} shows results for a non-diseased image.
 Figure~\ref{fig:seg1} (a) shows $I^{Flt}$ and Figure \ref{fig:seg1} (b) shows the superimposed segmentation masks of $I^{Ref}$ ($I^{Ref}_{Seg}$) and $I^{Flt}$ ($I^{Flt}_{Seg}$) on the reference image to give an idea of the degree of misalignment before registration. 
Figures~\ref{fig:seg1} (c)-(f) show the superimposed contours after registration using different methods. In these examples $I_{Seg}^{Trans}$ is shown in green with $I_{Seg}^{Ref}$ in red. $I_{Seg}^{Trans}$ is obtained by applying the recovered deformation field to $I_{Seg}^{Flt}$. Better registration is indicated by closer alignment of the two contours. Figure~\ref{fig:seg2} shows the corresponding results for the diseased case.

\begin{table}[t]
\begin{tabular}{|c|c|c|c|c|}
\hline
{} & \multicolumn {4}{|c|}{Normal Images}  \\  \cline{2-5} 
{} & {Bef.} & \multicolumn {3}{|c|}{After Registration}  \\ \cline{3-5} 
{} & {Reg} & {UReg-Net}  & {UReg$_{Image}$} & {DIRNet}  \\ \hline
%
{DM($\%$)} & {78.9} & {89.3}  & {85.2} & {84.8} \\ \hline
{HD$_{95}$(mm)} & {12.9} & {6.9}  & {8.4} & {8.7} \\ \hline
{MAD} & {13.7} & {7.3}  & {8.9} & {9.1} \\ \hline
{Time(s)} & {} & {0.5} & {0.4} & {0.6} \\ \hline
{} &  \multicolumn {4}{|c|}{After Registration}  \\ \cline{2-5} 
{} & {FlowNet} & {JRS-Net} & {Elastix} & {VoxelMorph} \\ \hline
%
{DM($\%$)} & {83.5} & {88.6} & {82.1} & {88.2} \\ \hline
{HD$_{95}$(mm)} & {9.8} & {7.5} & {10.8} & {7.4} \\ \hline
{MAD} & {13.7} & {8.6} & {11.1} & {7.9} \\ \hline
{Time(s)} & {0.5} & {0.6} & {21} & {0.6} \\ \hline
\end{tabular}
\caption{Image registration results for \textbf{left and right lung}  using different methods on \textbf{non-diseased images} of the NIH-14 database. $Time$ indicates computation time in seconds.}
\label{tab:NIH1}
\end{table}

\begin{table}[t]
\begin{tabular}{|c|c|c|c|c|}
\hline
{} & \multicolumn {4}{|c|}{Diseased Images} \\  \cline{2-5} 
{} & {Bef.} & \multicolumn {3}{|c|}{After Registration} \\ \cline{3-5} 
{} & {Reg} & {UReg-Net}  & {UReg$_{Image}$} & {DIRNet} \\ \hline
%
{DM($\%$)} & {79.1} & {88.9}  & {85.0} & {84.4} \\ \hline
{HD$_{95}$(mm)} & {11.8} & {7.3}  & {8.6} & {8.9} \\ \hline
{MAD} & {12.7} & {7.9}  & {9.2} & {9.6} \\ \hline
{Time(s)} & {} & {0.5} & {0.4} & {0.6} \\ \hline
{} & \multicolumn {4}{|c|}{After Registration} \\ \cline{2-5} 
{} & {FlowNet} & {JRS-Net} & {Elastix} & {Voxel Morph}\\ \hline
%
{DM($\%$)} & {83.1} & {88.2} & {81.5} & {87.9} \\ \hline
{HD$_{95}$(mm)} & {10.1} & {8.1} & {11.5} & {8.0} \\ \hline
{MAD} & {12.7} & {9.0} & {12.1} & {8.3} \\ \hline
{Time(s)} & {0.5} & {0.3} & {21} & {0.6} \\ \hline
\end{tabular}
\caption{Image registration results for \textbf{left and right lung} using different methods on diseased images of the NIH-14 database. $Time$ indicates computation time in seconds.}
\label{tab:NIH12}
\end{table}


\subsection{Accuracy of Generated Images}

Our method is different from previous image registration approaches since we directly generate the registered image instead of iteratively updating the image. Consequently, there is always a risk that the images generated by the GAN may not be realistic or the generated images are not diverse enough. To ensure that it is not the case we perform a classification experiment. 

We take a VGG-16 classifier \cite{VGG} pre-trained on the ImageNet dataset \cite{Imagenet} and use $700$ reference images, with data augmentation, to finetune it for a $14$ label classification problem which is the number of diseased classes represented in the NIH dataset. The finetuning is done under a $5-$ fold cross validation setting to get an average area under curve (AUC) of $82.5$ for the $14$ class disease classification problem. This value is the baseline against which other results are compared.

 After finetuning, the classifier is evaluated on the test set of $2112$ floating images and the corresponding registered images generated by the registration GAN. 
If the generated images are real then the classification performance on the generated images would be similar to the baseline.  The AUC values obtained for classifying the generated images was $81.9$ and $82.8$, respectively, for the real floating images and registered images generated by the GAN.  The \dm{two very similar values and their closeness to the baseline demonstrate that  the generated images have a high degree of realism.} 

\dm{
In another set of experiments we used a random noise vector instead of the latent vector from the auto encoder. The generated images are used to repeat the  classification experiment as described above. We obtain AUC values of $76.2$ with the VGG classifier, which is a significant decrease from the performance of latent vector encoding. The corresponding p-value from a t-test is $0.032$, indicating statistically different results and reinforce the fact that any arbitrary vector is not suitable for domain independent registration. Rather it is the use of CAEs that transforms the image to a latent feature space and exhibits enhanced performance through the generation of realistic images.}

\dm{
Furthermore, in another set of experiments we take the output of the CAEs and add small amounts of Gaussian noise (zero mean and $\sigma \in [0.2 1]$) to distort the latent feature vector. The corresponding classification experiment results in AUC values of $81.1$, with $p>0.0754$ compared to the previously mentioned baseline.  These results show our CAE based approach is fairly robust and, despite added noise, is able to generate realistic images.   
}

\dmt{
\subsection{Ablation Study Experiments}
}

Equations~\ref{eq:conLoss},\ref{eqn:adloss} denote the different terms used in the cost function. In this section we describe a set of experiments that evaluate the importance of different sources of information in achieving registration. Equation~\ref{eq:conLoss} has NMI, VGG and  SSIM terms. In one set of experiments we exclude the NMI term in the cost function and evaluate the images thus generated. The method is denoted as $UReg_{w-NMI}$, or $UReg-Net$ without NMI. We also generate results without VGG ($UReg_{w-VGG}$) and SSIM ($UReg_{w-SSIM}$) terms in Equation~\ref{eq:conLoss}, and by excluding   deformation field error, $\log (1-MSE_{Norm}(I^{Def-App},I^{Def-Recv})$), in Eqn.~\ref{eqn:adloss} ($UReg_{w-Def}$). 
%

Results of \dmt{ablation study experiments} on diseased and normal images are summarized in Tables~\ref{tab:NIH2},\ref{tab:NIH3}. The performance drops by a statistically significant measure when we exclude different terms of the cost function. 
Of particular importance are the numbers for $UReg_{Image}$ that show the maximum drop in performance compared to $UReg-Net$. This clearly demonstrates the advantage of using latent space representation instead of images for registration. 

\begin{table}[t] 
\begin{tabular}{|c|c|c|c|}
\hline
{} & {Bef.} & \multicolumn {2}{|c|}{After Registration}  \\ \cline{3-4} 
{} & {Reg}  & {UReg$_{Image}$} & {UReg$_{w-NMI}$}  \\ \hline
{DM($\%$)} & {78.9}  & {85.2} & {83.4} \\ \hline
{HD$_{95}$(mm)} & {12.9}  & {8.4} & {8.9} \\ \hline
{MAD(mm)} & {13.7}  & {8.9} & {9.5} \\ \hline
{Time(s)} & {} & {0.4} & {0.4} \\ \hline
{} & \multicolumn {3}{|c|}{After Registration}  \\ \cline{2-4} 
{} & {UReg$_{w-VGG}$} & {UReg$_{w-SSIM}$} & {UReg$_{w-Def}$} \\ \hline
{DM($\%$)} & {81.9} & {82.6} & {82.1} \\ \hline
{HD$_{95}$(mm)} & {9.9} & {9.2} & {10.1} \\ \hline
{MAD(mm)} & {13.7}  & {9.8} & {10.9} \\ \hline
{Time(s)} & {0.4} & {0.4} & {0.4} \\ \hline
\end{tabular}
\caption{Image registration results for normal images by excluding different terms in the cost function. $Time$ indicates computation time in seconds.}
\label{tab:NIH2}
\end{table}

\begin{table}[t] 
\begin{tabular}{|c|c|c|c|}
\hline
{} & {Bef.} & \multicolumn {2}{|c|}{After Registration} \\ \cline{3-4} 
{} & {Reg}  & {UReg$_{Image}$} & {UReg$_{w-NMI}$}  \\ \hline
{DM($\%$)} & {79.1}  & {85.0} & {83.2} \\ \hline
{HD$_{95}$(mm)} & {11.8}  & {8.6} & {9.3} \\ \hline
{MAD(mm)} & {12.7}  & {9.2} & {9.8} \\ \hline
{Time(s)} & {} & {0.4} & {0.4} \\ \hline
{} &  \multicolumn {3}{|c|}{After Registration} \\ \cline{2-4} 
{} & {UReg$_{w-VGG}$} & {UReg$_{w-SSIM}$} & {UReg$_{w-Def}$} \\ \hline
{DM($\%$)} & {82.0} & {82.3} & {81.8} \\ \hline
{HD$_{95}$(mm)} & {10.4} & {9.1} & {10.3} \\ \hline
{MAD(mm)} & {11.2} & {10.1} & {11.1} \\ \hline
{Time(s)} & {0.4} & {0.4} & {0.4} \\ \hline
\end{tabular}
\caption{Image registration results for diseased images by excluding different terms in the cost function. $Time$ indicates computation time in seconds.}
\label{tab:NIH3}
\end{table}

\subsection{Validation of Synthetic Images By Clinical Experts}

\dm{
We also validate the authenticity of generated images through visual examination by clinical experts.
A $90$ minutes session was set with three experienced radiologists each having over $15$ years of experience in the diagnosis of lung conditions from X-ray/CT images. They are tasked with providing feedback whether the generated registered images and their deformation fields are realistic or not. $150$ reference and floating image pairs are selected, and registered images registered using three different approaches: 1) conventional image registration using Elastix; 2) GAN based registration using the original images ($GAN_{Image}$); and 3) our GAN based registration method using latent vector representations of the images as input ($GAN_{Z}$). For 2) we re-trained our network, replacing $z$ with images.   
The selected images had a uniform distribution amongst the $14$ disease classes. For each case the $I^{Ref}$ and $I^{Flt}$ were first displayed to all three radiologists, followed by the display of $I^{Trans}$ and $I^{Def-Recv}$. The experts were asked to indicate whether $I^{Trans}$ is realistic or not, and were agnostic to the method that produced the registered images. Each expert viewed all registered images obtained by each of the three methods. 
}

\dm{
The results for consensus on realism is summarized in Table~\ref{tab:realism}. The results of Elastix based registration act as a baseline and, expectedly, have high agreement on realism since there are no generative models involved. The few images that are deemed as unrealistic are, possibly, a result of inaccurate registration. For $GAN_{Image}$ the agreement drops by about $4-5\%$. This is not unexpected and has to do with the capacity of the implemented GAN model for learning desired image distributions. In the case of $GAN_Z$ the agreement drops slightly by $1-1.5\%$. Although it indicates a reduced ability to generate realisitc images, the difference with respect to $GAN_{Image}$ is not significant as indicated by a p-value of $0.0867$. Thus we can conclude that use of latent vector representations does not significantly degrade the realism of generated images. But it provides the advantage of using the registration method for multiple problem domains without laborious retraining. 
}

\begin{table}[t] 
\begin{tabular}{|c|c|c|c|}
\hline
{Agreement} & \multicolumn {3}{|c|}{Registration Method} \\ \cline{2-4} 
{Statistics} & {Elastix}  & {$GAN_{Image}$} & {$GAN_Z$}  \\ \hline
{All 3 Experts} & {\textbf{91.3}~(137)}  & {\textbf{84.7}~(127)} & {\textbf{83.4}~(125)} \\ \hline
{Atleast 2 Experts} & {\textbf{94.0}~(141)}  & {\textbf{87.3}~(131)} & {\textbf{85.3}~(128)} \\ \hline
{Atleast 1 Expert} & {\textbf{98.7}~(148)}  & {\textbf{89.3}~(134)} & {\textbf{87.3}~(131)} \\ \hline
{No Experts} & {\textbf{1.3}~(2)} & {\textbf{10.7}~(16)} & {\textbf{12.7}~(19)} \\ \hline
\end{tabular}
\caption{Agreement statistics for different registration methods amongst 3 radiologists. Numbers in bold indicate percentage of agreement among clincians while numbers within brackets indicate actual numbers out of $150$ patients.}
\label{tab:realism}
\end{table}

\subsection{Atlas Registration Results Using Brain MRI}
\label{expt:atlas}

We demonstrate our method on the task of brain MRI registration. We used the $800$ images of the ADNI-1 dataset \cite{Bala33} consisting of $200$ controls, $400$ MCI and $200$ Alzheimer's Disease patients. The MRI protocol for ADNI1 focused on consistent longitudinal structural imaging on $1.5T$ scanners using $T1$ and dual echo $T2-$weighted sequences. 
%
All scans were resampled to $256\times256\times256$ with $1$mm isotropic voxels. Pre-processing includes affine registration and brain extraction using FreeSurfer \cite{Bala17}, and cropping the resulting images to $160 \times 192 \times 224$. 
%
%
The dataset is split into $560$, $120$, and $120$ volumes for training, validation, and testing. We simulate elastic deformations to generate the floating images  using which we evaluate registration performance. Registration performance was calculated for $2$D slices. 
%

\begin{figure}[t]
\begin{tabular}{ccc}
\includegraphics[height=2.5cm,width=2.5cm]{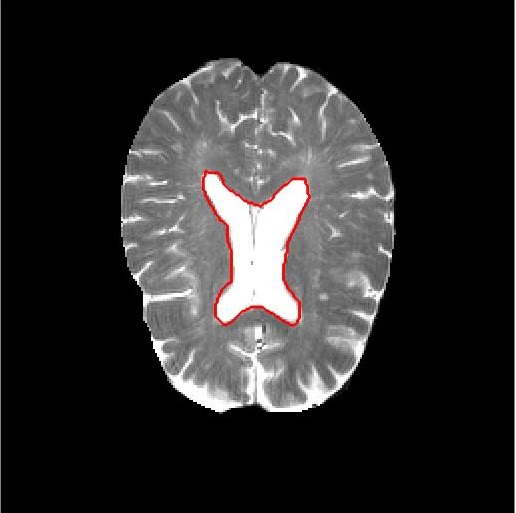} &
\includegraphics[height=2.5cm,width=2.5cm]{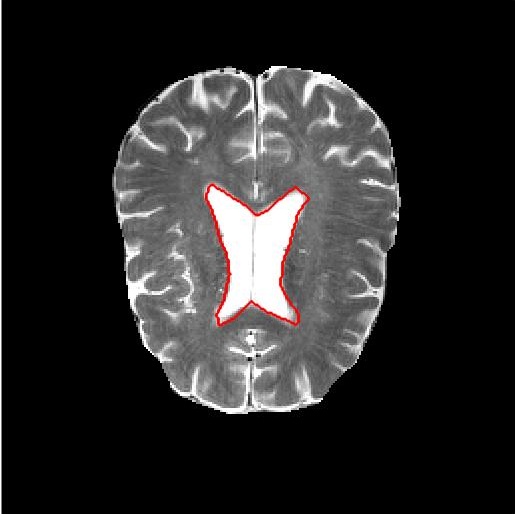} &
\includegraphics[height=2.5cm,width=2.5cm]{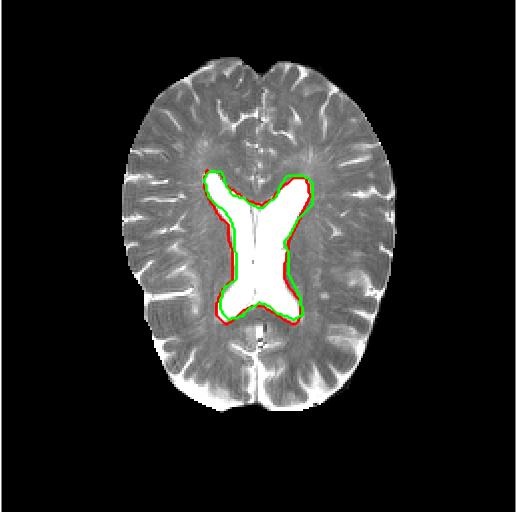} \\
(a) & (b) & (c)  \\
\includegraphics[height=2.5cm,width=2.5cm]{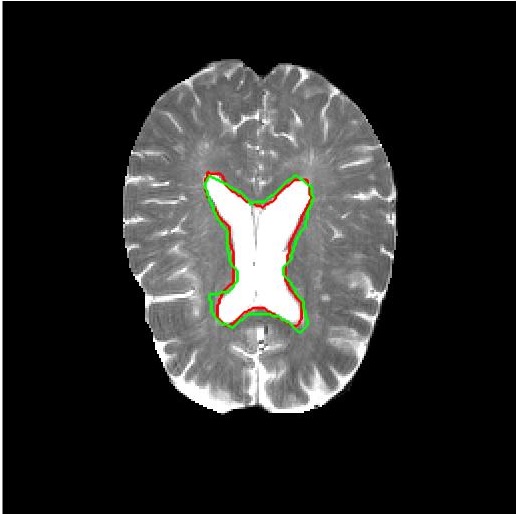} &
\includegraphics[height=2.5cm,width=2.5cm]{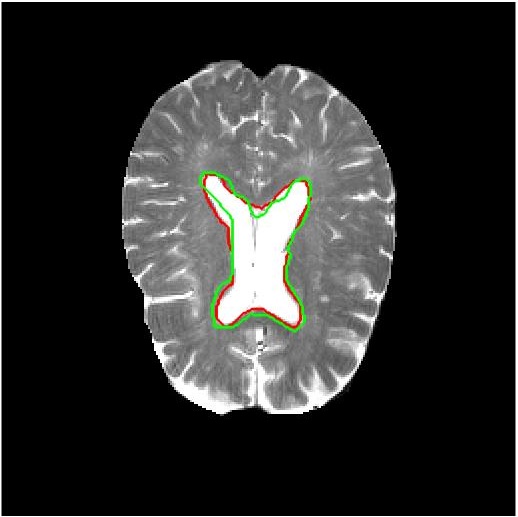} &
\includegraphics[height=2.5cm,width=2.5cm]{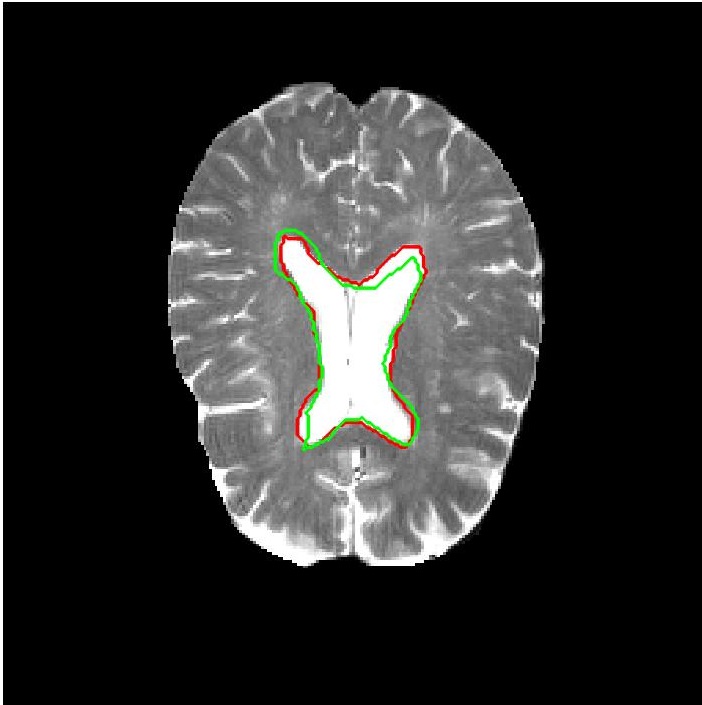} \\
(d) & (e) & (f) \\ 
\includegraphics[height=2.5cm,width=2.5cm]{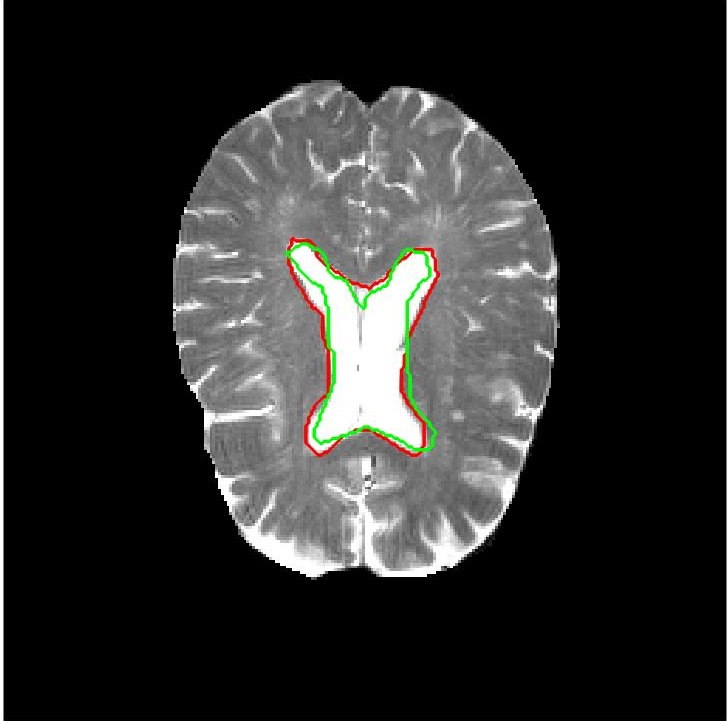} &
\includegraphics[height=2.5cm,width=2.5cm]{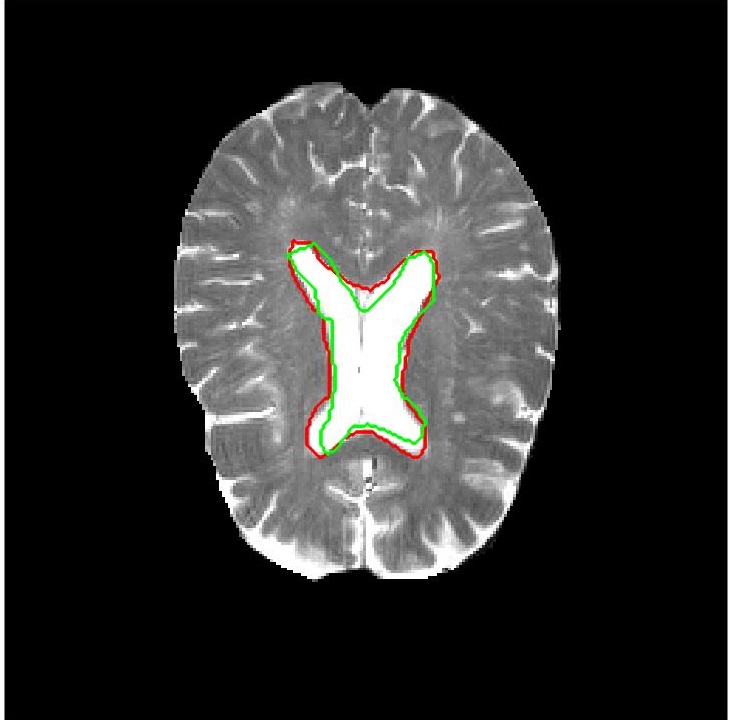} &
\includegraphics[height=2.5cm,width=2.5cm]{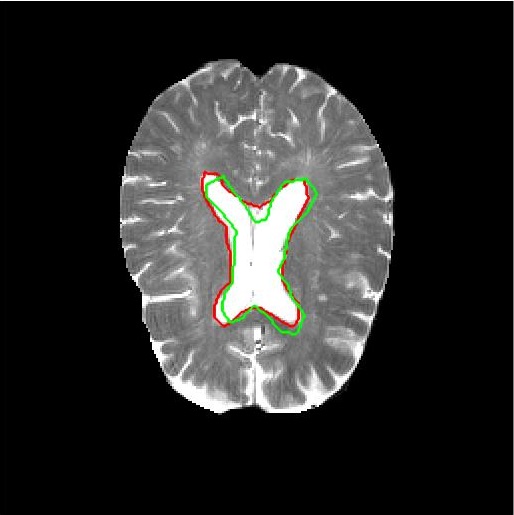} \\
(g) & (h) & (i) \\
\end{tabular}
\caption{Results for atlas based brain MRI image registration. (a) $I_{Ref}$ with $I_{Ref}^{Seg}$ (b) $I_{Flt}$ with $I_{Flt}^{Seg}$. Superimposed registered mask (in green) obtained using: (c) $UReg-Net$; (d)\cite{BalaCVPR18}; (e) $UReg_{Image}$; (f) $DIR-Net$; (g) $JRS-Net$; (f) \cite{FlowNet} and (g) $Elastix$.}
\label{fig:brain}
\end{figure}

Figure~\ref{fig:brain} shows results for image registration. We show the reference image (or the atlas image) in  Figure~\ref{fig:brain} (a) followed by an example floating image in Figure~\ref{fig:brain} (b). The ventricle structure to be aligned is shown in red in both images. Similar to the results of lung images, Figure~\ref{fig:brain} (c)-(i) show the deformed  structures obtained by applying the registration field obtained from different methods to the floating image and superimposing these structures on the atlas image. The deformed structures from the floating image are shown in green. In case of a perfect registration the green and red contours should coincide perfectly. 
Table~\ref{tab:brain} shows the results of atlas image based brain registration using the average Dice scores between structures before and after registration. 

\begin{table}[t]
\begin{tabular}{|c|c|c|c|c|}
\hline
{} & {Bef.} & \multicolumn {3}{|c|}{After Registration}  \\ \cline{3-5} 
{} & {Reg} & {URegNet}  & {UReg$_{Image}$} & {DIRNet}   \\ \hline
%
{DM($\%$)} & {67.2} & {78.5}  & {74.6} & {75.1} \\ \hline
{HD$_{95}$(mm)} & {14.5} & {9.4}  & {11.1} & {11.4} \\ \hline
{MAD} & {16.1} & {10.5}  & {12.1} & {12.7} \\ \hline
{Time(s)} & {} & {0.5} & {0.4} & {0.6} \\ \hline
{} & \multicolumn {4}{|c|}{After Registration}  \\ \cline{2-5} 
{} & {FlowNet} & {JRS-Net} & {Elastix} & {VoxelMorph} \\ \hline 
%
{DM($\%$)} & {73.2} & {76.8} & {73.1} & {74.9} \\ \hline
{HD$_{95}$(mm)} & {12.6} & {11.8} & {13.8} & {11.9} \\ \hline
{MAD} & {13.9} & {13.1} & {15.0} & {12.8} \\ \hline
{Time(s)} & {0.5} & {0.6} & {21} & {0.5} \\ \hline
\end{tabular}
\caption{Atlas based registration results for different methods on brain images. }
\label{tab:brain}
\end{table}

\subsection{Performance on New Test Images}

So far we have shown results where the training and test set were of the same anatomies (i.e., lung to lung and brain to brain). In this section we show results for cases where the training was performed with one image type and used to register images of the other type. Table~\ref{tab:newtest1} summarizes the performance of all the deep learning networks when trained on the diseased lung images and used to register brain images, while Table~\ref{tab:newtest2} summarizes performance when the networks were trained on brain images and used to register diseased lung images. Compared to the results for lung (Table~\ref{tab:NIH12}) and brain (Table~\ref{tab:brain}) images, it is obvious that for the same type of training and test images registration performance is much better than having different training and test images. Despite these limitations our proposed method, $UReg-Net$, performs better than other methods and the values are close to the case of having same train and test image.
Note that the Elastix method does not involve learning and acts as a baseline, while other methods are trained on images different from the test images

\begin{table}[t]
\begin{tabular}{|c|c|c|c|c|}
\hline
{} & {Bef.} & \multicolumn {3}{|c|}{After Registration}  \\ \cline{3-5} 
{} & {Reg} & {UReg-Net}  & {UReg$_{Image}$} & {DIRNet}   \\ \hline
%
{DM($\%$)} & {67.2} & {74.1}  & {71.2} & {70.9} \\ \hline
{HD$_{95}$(mm)} & {14.5} & {10.8}  & {12.3} & {12.7} \\ \hline
{MAD} & {16.1} & {12.0}  & {13.6} & {14.1} \\ \hline
{Time(s)} & {} & {0.5} & {0.4} & {0.6} \\ \hline
{} & \multicolumn {4}{|c|}{After Registration}  \\ \cline{2-5} 
{} & {FlowNet} & {JRS-Net} & {Elastix} & {VoxelMorph} \\ 
%
{DM($\%$)} & {69.4} & {72.2} & {71.1} & {71.3} \\ \hline
{HD$_{95}$(mm)} & {13.1} & {13.8} & {14.2} & {13.6} \\ \hline
{MAD} & {15.1} & {14.6} & {15.9} & {14.2} \\ \hline
{Time(s)} & {0.5} & {0.6} & {21} & {0.5} \\ \hline
\end{tabular}
\caption{Registration results for brain images when network is trained on diseased lung images. }
\label{tab:newtest1}
\end{table}

\begin{table}[t]
\begin{tabular}{|c|c|c|c|c|}
\hline
{} & \multicolumn {4}{|c|}{Diseased Images} \\  \hline
{} & {Bef.} & \multicolumn {3}{|c|}{After Registration} \\ \cline{3-5} 
{} & {Reg} & {UReg-Net}  & {UReg$_{Image}$} & {DIRNet} \\  \hline
%
{DM($\%$)} & {79.1} & {84.6}  & {82.0} & {82.1} \\ \hline
{HD$_{95}$(mm)} & {11.8} & {8.8}  & {10.0} & {10.3} \\ \hline
{MAD} & {12.7} & {9.5}  & {10.8} & {10.9} \\ \hline
{Time(s)} & {} & {0.5} & {0.4} & {0.6} \\ \hline
{} & \multicolumn {4}{|c|}{Diseased Images} \\  \hline
{} & \multicolumn {4}{|c|}{After Registration} \\ \cline{2-5} 
{} & {FlowNet} & {JRS-Net} & {Elastix} & {VoxelMorph}\\ \hline 
%
{DM($\%$)} & {81.7} & {83.1} & {81.6} & {82.5} \\ \hline
{HD$_{95}$(mm)} & {9.9} & {11.1} & {11.6} & {10.2} \\ \hline
{MAD} & {12.3} & {11.6} & {13.1} & {11.6} \\ \hline
{Time(s)} & {0.5} & {0.4} & {21} & {0.6} \\ \hline
\end{tabular}
\caption{Registration results for diseased lung images when network is trained on brain images.}
\label{tab:newtest2}
\end{table}

\subsection{Retinal Image Registration Results}
\label{expt:retina}

The dataset consists of retinal colour fundus images and fluorescein angiography (FA) images obtained from $30$ normal subjects. Both images are $576\times720$ pixels and fovea centred \cite{Alipour2014}. 
Registration ground truth was developed using the Insight Toolkit (ITK). 
The Frangi vesselness\cite{Frangi1998} feature was utilised to find the vasculature, and the maps were aligned using sum of squared differences (SSD). 
Three out of $30$ images could not be aligned due to poor contrast and one FA image was missing, leaving us with a final set of $26$ registered pairs. 
We use the fundus images as $I^{Ref}$ and generate floating images from the FA images by simulating different deformations (using SimpleITK) such as rigid, affine and elastic deformations(maximum displacement of a pixel was $\pm10$ mm. $1500$ sets of deformations were generated for each image pair giving a total of $39000$ image pairs to be evaluated for registration performance.

Our algorithm's performance was evaluated using  previously mentioned metrics.
Before applying simulated deformation the mean Dice overlap of the vasculature between the fundus and FA images across all $26$ patients is $99.2$, which indicates highly accurate alignment. After simulating deformations the individual Dice overlap reduces considerably depending upon the extent of deformation. The Dice value after successful registration is expected to be higher than before registration. We also calculate the $95$ percentile  Hausdorff Distance ($HD_{95}$) and the mean absolute surface distance (MAD) before and after registration. We calculate the mean square error (MSE) between the registered FA image and the original undeformed FA image to quantify their similarity. 
The intensity of both images was normalized to lie in $[0,1]$. 

\begin{figure*}[t]
\begin{tabular}{cccc}
\includegraphics[height=2.2cm,width=2.2cm]{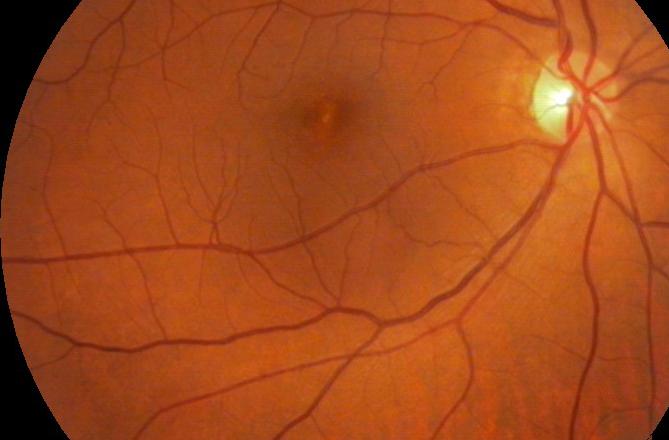} &
\includegraphics[height=2.2cm,width=2.2cm]{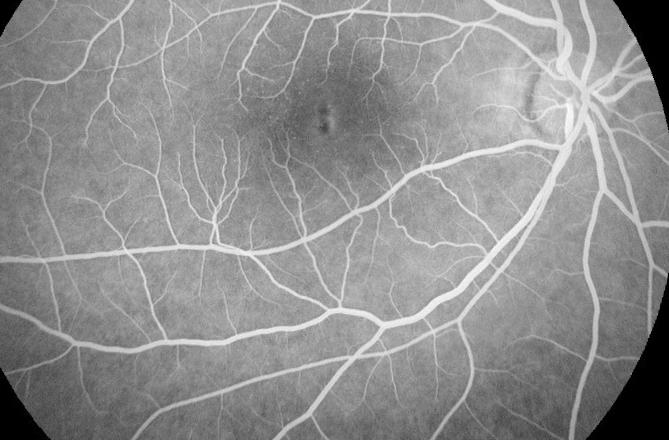} &
\includegraphics[height=2.2cm,width=2.2cm]{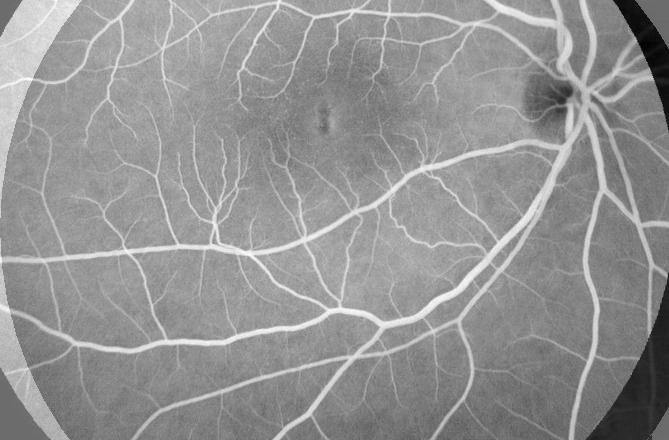} &
\includegraphics[height=2.2cm,width=2.2cm]{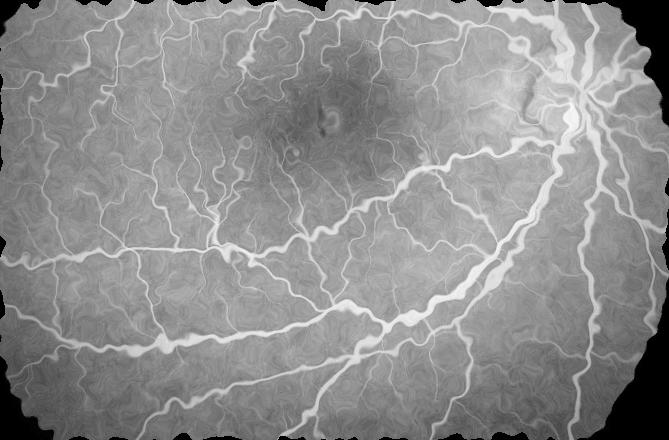} \\
(a) & (b) & (c) & (d) \\
\includegraphics[height=2.2cm,width=2.2cm]{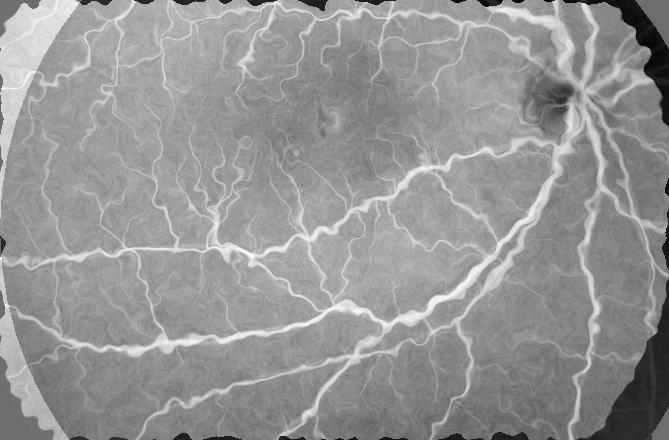} &
\includegraphics[height=2.2cm,width=2.2cm]{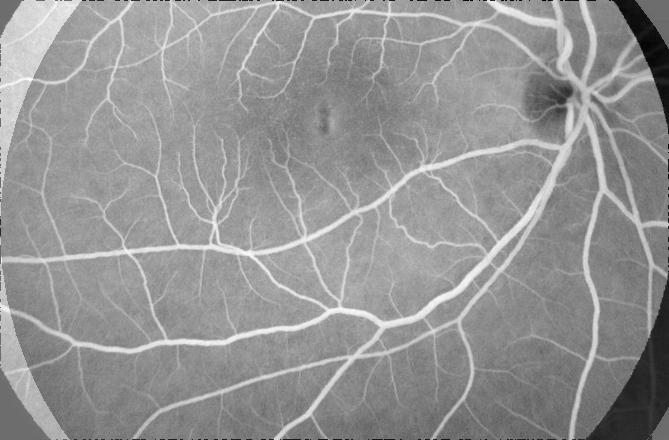} &
\includegraphics[height=2.2cm,width=2.2cm]{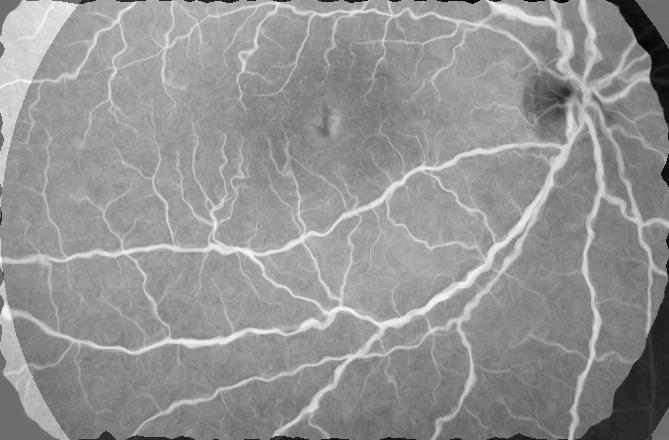} &
\includegraphics[height=2.2cm,width=2.2cm]{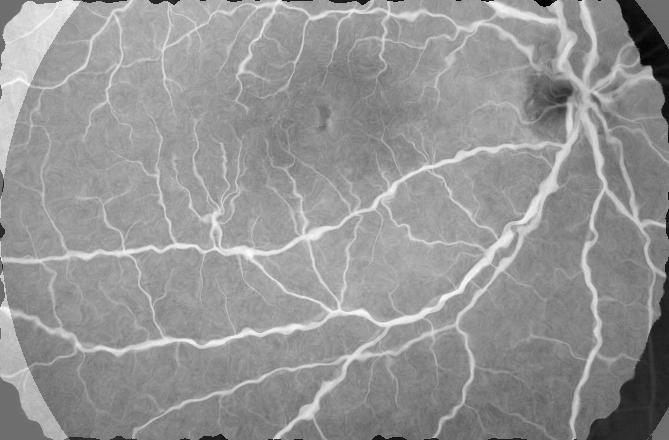} \\
 (e) & (f) & (g) & (h)\\ 
\end{tabular}
\caption{Example results for retinal fundus and FA registration. (a) Color fundus image; (b) Original FA image; (c) ground truth difference image before simulated deformation; (d) Deformed FA image or the floating image; Difference image (e) before registration; after registration using (f) $UReg-Net$; (g) $DIRNet$; (h) Elastix .}
\label{fig:Ret}
\end{figure*}

In this scenario all the deep learning networks were trained on the chest xray images and used to register the retinal image pairs.
Table~\ref{tab:Retina} shows the registration performance for the different methods. Figure~\ref{fig:Ret} shows registration results for retinal images. It is  observed (for retinal, lung and brain image registration) that the deep learning methods perform better than Elastix for same training and test images but are not significantly different for different train and test pairs, except for $UReg-Net$ which is designed to overcome this challenge. Thus our approach of using latent space feature maps is highly effective in achieving accurate registration on new test images.

\begin{table}[t]
\begin{tabular}{|c|c|c|c|c|}
\hline
{} & {Bef.} & \multicolumn {3}{|c|}{After Registration} \\ \cline{3-5} 
{} & {Reg} & {UReg-Net}  & {UReg$_{Image}$} & {DIRNet} \\  \hline
%
{DM($\%$)} & {72.5} & {83.9}  & {81.8} & {81.3} \\ \hline
{HD$_{95}$(mm)} & {12.7} & {9.6}  & {10.9} & {11.2} \\ \hline
{MAD} & {13.4} & {10.2}  & {11.6} & {11.8} \\ \hline
{Time(s)} & {} & {0.3} & {0.3} & {0.4} \\ \hline
{} & \multicolumn {4}{|c|}{After Registration} \\ \cline{2-5} 
{} & {FlowNet} & {JRS-Net} & {Elastix} & {VoxelMorph}\\ \hline 
%
{DM($\%$)} & {81.1} & {82.1} & {81.1} & {81.7} \\ \hline
{HD$_{95}$(mm)} & {10.7} & {11.9} & {12.5} & {10.9} \\ \hline
{MAD} & {13.1} & {12.4} & {14.1} & {12.2} \\ \hline
{Time(s)} & {0.3} & {0.6} & {18} & {0.4} \\ \hline
\end{tabular}
\caption{Registration results for retinal images when network is trained on chest xray images.}
\label{tab:Retina}
\end{table}

%
\section{Discussions}
\label{sec:disc}

\dm{
The images generated by our method exhibit a high degree of realism. Although GANs have their limitations in image generation, our approach produces satisfactory results in most cases as supported by the agreement statistics in Table~\ref{tab:realism}. We hypothesize that imposing very specific constraints during model training by adding cost terms, although narrowing the search space, may contribute to more realistic images since the network is trained to generate images satisfying very specific criteria. However this is realizable only in the presence of sufficient training data. In future work we aim to investigate this hypothesis which could lead to a better understanding of GAN training.  
}

\dm{
The latent space representation contributes towards more accurate registration by learning the most effective representation of the image. Medical images have lower dimension and resolution than natural images and hence provide less information. Consequently a lower dimensional representation, as in the case of latent space representation, contributes in improving registration performance by guiding the network to focus on most informative and discriminative regions. If we add another convolution and max pooling layer such that the latent feature map is $32\times32$ we observe a drop in registration accuracy. On the other hand if the latent feature map is $128\times128$, there is no significant improvement in registration performance but increases the training time by nearly $6$ times. Hence we choose to use the $64\times64$ maps as the optimal representation.
}

\dm{
While our method does improve on existing methods, we also observe a few failure cases. One common scenario is when the floating image has multiple disease labels. Although our method is not dependent on the disease label, multiple labels indicate presence of multiple pathologies and hence the images are hazy in appearance. Consequently, the registration output is not accurate as multiple pathologies distort the discriminative information that can be extracted from the image. In such scenarios using the original images instead of latent feature maps does not make much difference to the result. 
}
\dm{
Although we demonstrate that the network trained on Xray images can register retinal images, the results can be much improved if the training and test imaging modality are similar. These observations motivate us to explore other image feature representation methods such that the common feature space is equally informative and discriminative irrespective of the input images. 
}

\section{Conclusion}
\label{sec:concl}

We have proposed a novel deep learning framework for registration of different types of medical images using unsupervised domain adaptation and generative adversarial networks. Our proposed method achieves dataset independent registration where it is trained on one kind of images and achieves state of the art performance in registering different image types. GANs are trained to generate the registered image and the corresponding deformation field. The primary novelty of our method is the use of latent space feature maps from autoenoders that facilitates dataset independent registration. 
Experimental results show our domain adaptation based registration method performs better than existing methods that rely on vast amounts of training data for image registration.
\dm{On the one hand CAE feature maps lead to better registration by reducing the input feature dimension, while on the other hand they also open up other research questions related to GAN training and image feature representation. We hope to address these questions in future work.
}


\bibliographystyle{elsarticle-num}
\bibliography{PR_SI_2019_Ref}

\begin{thebibliography}{100}
\expandafter\ifx\csname url\endcsname\relax
  \def\url#1{\texttt{#1}}\fi
\expandafter\ifx\csname urlprefix\endcsname\relax\def\urlprefix{URL }\fi
\expandafter\ifx\csname href\endcsname\relax
  \def\href#1#2{#2} \def\path#1{#1}\fi

\bibitem{Mahapatra_CVIU2019}
B.~Bozorgtabar, D.~Mahapatra, H.~von Teng, A.~Pollinger, L.~Ebner, J.-P.
  Thiran, M.~Reyes, Informative sample generation using class aware generative
  adversarial networks for classification of chest xrays., Computer Vision and
  Image Understanding 184 (2019) 57--65 (2019).

\bibitem{Mahapatra_PR2020}
D.~Mahapatra, Z.~Ge, Training data independent image registration using
  generative adversarial networks and domain adaptation., In press Pattern
  Recognition 100 (2020) 1--14 (2020).

\bibitem{Mahapatra_CMIG2019}
D.~Mahapatra, B.~Bozorgtabar, R.~Garnavi, Image super-resolution using
  progressive generative adversarial networks for medical image analysis.,
  Computerized Medical Imaging and Graphics 71 (2019) 30--39 (2019).

\bibitem{Mahapatra_LME_PR2017}
D.~Mahapatra, Semi-supervised learning and graph cuts for consensus based
  medical image segmentation., Pattern Recognition 63~(1) (2017) 700--709
  (2017).

\bibitem{Zilly_CMIG_2016}
J.~Zilly, J.~Buhmann, D.~Mahapatra, Glaucoma detection using entropy sampling
  and ensemble learning for automatic optic cup and disc segmentation., In
  Press Computerized Medical Imaging and Graphics 55~(1) (2017) 28--41 (2017).

\bibitem{BalaCVPR18}
G.~Balakrishnan, A.~Zhao, M.~Sabuncu, J.~Guttag, An supervised learning model
  for deformable medical image registration, in: Proc. CVPR, 2018, pp.
  9252--9260 (2018).

\bibitem{RegNet}
H.~Sokooti, B.~de~Vos, F.~Berendsen, B.~Lelieveldt, I.~Isgum, M.~Staring,
  Nonrigid image registration using multiscale 3d convolutional neural
  networks, in: MICCAI, 2017, pp. 232--239 (2017).

\bibitem{Mahapatra_MLMI2018}
D.~Mahapatra, Z.~Ge, S.~Sedai, R.~Chakravorty, Joint registration and
  segmentation of xray images using generative adversarial networks, in:
  MICCAI-MLMI, 2018, pp. 73--80 (2018).

\bibitem{ZGe_MTA2019}
Z.~Ge, D.~Mahapatra, X.~Chang, Z.~Chen, L.~Chi, H.~Lu, Improving multi-label
  chest x-ray disease diagnosis by exploiting disease and health labels
  dependencies., In press Multimedia Tools and Application (2019) 1--14 (2019).

\bibitem{Mahapatra_SSLAL_CD_CMPB}
D.~Mahapatra, F.~Vos, J.~Buhmann, Active learning based segmentation of crohns
  disease from abdominal mri., Computer Methods and Programs in Biomedicine
  128~(1) (2016) 75--85 (2016).

\bibitem{Mahapatra_SSLAL_Pro_JMI}
D.~Mahapatra, J.~Buhmann, Visual saliency based active learning for prostate
  mri segmentation., SPIE Journal of Medical Imaging 3~(1) (2016).

\bibitem{Mahapatra_LME_CVIU}
D.~Mahapatra, Combining multiple expert annotations using semi-supervised
  learning and graph cuts for medical image segmentation., Computer Vision and
  Image Understanding 151~(1) (2016) 114--123 (2016).

\bibitem{LiTMI_2015}
Z.~Li, D.~Mahapatra, J.Tielbeek, J.~Stoker, L.~van Vliet, F.~Vos, Image
  registration based on autocorrelation of local structure., IEEE Trans. Med.
  Imaging 35~(1) (2016) 63--75 (2016).

\bibitem{PR_Falez}
P.~Falez, P.~Tirilly, I.~Bilasco, P.~Devienne, P.~Boulet, Unsupervised visual
  feature learning with spike-timing-dependent plasticity: {H}ow far are we
  from traditional feature learning approaches?., Pattern Recognition 93 (2019)
  418--429 (2019).

\bibitem{MahapatraJDI_Cardiac_FSL}
D.~Mahapatra, Automatic cardiac segmentation using semantic information from
  random forests., J. Digit. Imaging. 27~(6) (2014) 794--804 (2014).

\bibitem{Mahapatra_JSTSP2014}
D.~Mahapatra, S.~Gilani, M.~Saini., Coherency based spatio-temporal saliency
  detection for video object segmentation., IEEE Journal of Selected Topics in
  Signal Processing. 8~(3) (2014) 454--462 (2014).

\bibitem{Behzad_PR2020}
B.~Bozorgtabar, D.~Mahapatra, J.-P. Thiran, Exprada: Adversarial domain
  adaptation for facial expression analysis., In Press Pattern Recognition 100
  (2020) 15--28 (2020).

\bibitem{MahapatraTIP_RF2014}
D.~Mahapatra, J.~Buhmann, Analyzing training information from random forests
  for improved image segmentation., IEEE Trans. Imag. Proc. 23~(4) (2014)
  1504--1512 (2014).

\bibitem{MahapatraTBME_Pro2014}
D.~Mahapatra, J.~Buhmann, Prostate mri segmentation using learned semantic
  knowledge and graph cuts., IEEE Trans. Biomed. Engg. 61~(3) (2014) 756--764
  (2014).

\bibitem{RegRev}
J.~maintz, M.~Viergever, A survey of medical image registration, Med. Imag.
  Anal 2~(1) (1998) 1--36 (1998).

\bibitem{MahapatraTMI_CD2013}
D.~Mahapatra, J.Tielbeek, J.~Makanyanga, J.~Stoker, S.~Taylor, F.~Vos,
  J.~Buhmann, Automatic detection and segmentation of crohn's disease tissues
  from abdominal mri., IEEE Trans. Med. Imaging 32~(12) (2013) 1232--1248
  (2013).

\bibitem{MahapatraJDICD2013}
D.~Mahapatra, J.Tielbeek, F.~Vos, J.~Buhmann, A supervised learning approach
  for crohn's disease detection using higher order image statistics and a novel
  shape asymmetry measure., J. Digit. Imaging 26~(5) (2013) 920--931 (2013).

\bibitem{MahapatraJDIMutCont2013}
D.~Mahapatra, Cardiac mri segmentation using mutual context information from
  left and right ventricle., J. Digit. Imaging 26~(5) (2013) 898--908 (2013).

\bibitem{MahapatraJDIGCSP2013}
D.~Mahapatra, Cardiac image segmentation from cine cardiac mri using graph cuts
  and shape priors., J. Digit. Imaging 26~(4) (2013) 721--730 (2013).

\bibitem{MahapatraJDIJSGR2013}
D.~Mahapatra, Joint segmentation and groupwise registration of cardiac
  perfusion images using temporal information., J. Digit. Imaging 26~(2) (2013)
  173--182 (2013).

\bibitem{WuTBME}
G.~Wu, M.~Kim, Q.~Wang, B.~C. Munsell, , D.~Shen., Scalable high performance
  image registration framework by unsupervised deep feature representations
  learning., IEEE Trans. Biomed. Engg. 63~(7) (2016) 1505--1516 (2016).

\bibitem{MahapatraJDISkull2012}
D.~Mahapatra, Skull stripping of neonatal brain mri: Using prior shape
  information with graphcuts., J. Digit. Imaging 25~(6) (2012) 802--814 (2012).

\bibitem{MahapatraTIP2012}
D.~Mahapatra, Y.~Sun, Integrating segmentation information for improved
  mrf-based elastic image registration., IEEE Trans. Imag. Proc. 21~(1) (2012)
  170--183 (2012).

\bibitem{MahapatraTBME2011}
D.~Mahapatra, Y.~Sun, Mrf based intensity invariant elastic registration of
  cardiac perfusion images using saliency information, IEEE Trans. Biomed.
  Engg. 58~(4) (2011) 991--1000 (2011).

\bibitem{MahapatraEURASIP2010}
D.~Mahapatra, Y.~Sun, Rigid registration of renal perfusion images using a
  neurobiology based visual saliency model, EURASIP Journal on Image and Video
  Processing. (2010) 1--16 (2010).

\bibitem{Miao_Reg}
S.~Miao, Y.~Z. Z.J.~Wang, R.~Liao, Real-time 2d/3d registration via cnn
  regression, in: IEEE ISBI, 2016, pp. 1430--1434 (2016).

\bibitem{Kuanar_ICIP19}
S.~Kuanar, V.~Athitsos, D.~Mahapatra, K.~Rao, Z.~Akhtar, D.~Dasgupta, Low dose
  abdominal ct image reconstruction: An unsupervised learning based approach,
  in: In Proc. IEEE ICIP, 2019, pp. 1351--1355 (2019).

\bibitem{Bozorgtabar_ICCV19}
B.~Bozorgtabar, M.~S. Rad, D.~Mahapatra, J.-P. Thiran, Syndemo: Synergistic
  deep feature alignment for joint learning of depth and ego-motion, in: In
  Proc. IEEE ICCV, 2019 (2019).

\bibitem{Xing_MICCAI19}
Y.~Xing, Z.~Ge, R.~Zeng, D.~Mahapatra, J.~Seah, M.~Law, T.~Drummond,
  Adversarial pulmonary pathology translation for pairwise chest x-ray data
  augmentation, in: In Proc. MICCAI, 2019, pp. 757--765 (2019).

\bibitem{Mahapatra_ISBI19}
D.~Mahapatra, Z.~Ge, Training data independent image registration with gans
  using transfer learning and segmentation information, in: In Proc. IEEE ISBI,
  2019, pp. 709--713 (2019).

\bibitem{MahapatraAL_MICCAI18}
D.~Mahapatra, S.~Bozorgtabar, J.-P. Thiran, M.~Reyes, Efficient active learning
  for image classification and segmentation using a sample selection and
  conditional generative adversarial network, in: In Proc. MICCAI (2), 2018,
  pp. 580--588 (2018).

\bibitem{Liao_Reg}
R.~Liao, S.~Miao, P.~de~Tournemire, S.~Grbic, A.~Kamen, T.~Mansi, D.~Comaniciu,
  An artificial agent for robust image registration, in: AAAI, 2017, pp.
  4168--4175 (2017).

\bibitem{Mahapatra_MLMI18}
D.~Mahapatra, Z.~Ge, S.~Sedai, R.~Chakravorty., Joint registration and
  segmentation of xray images using generative adversarial networks, in: In
  Proc. MICCAI-MLMI, 2018, pp. 73--80 (2018).

\bibitem{Sedai_OMIA18}
S.~Sedai, D.~Mahapatra, B.~Antony, R.~Garnavi, Joint segmentation and
  uncertainty visualization of retinal layers in optical coherence tomography
  images using bayesian deep learning, in: In Proc. MICCAI-OMIA, 2018, pp.
  219--227 (2018).

\bibitem{Sedai_MLMI18}
S.~Sedai, D.~Mahapatra, Z.~Ge, R.~Chakravorty, R.~Garnavi, Deep multiscale
  convolutional feature learning for weakly supervised localization of chest
  pathologies in x-ray images, in: In Proc. MICCAI-MLMI, 2018, pp. 267--275
  (2018).

\bibitem{MahapatraGAN_ISBI18}
D.~Mahapatra, B.~Antony, S.~Sedai, R.~Garnavi, Deformable medical image
  registration using generative adversarial networks, in: In Proc. IEEE ISBI,
  2018, pp. 1449--1453 (2018).

\bibitem{Sedai_MICCAI17}
S.~Sedai, D.~Mahapatra, S.~Hewavitharanage, S.~Maetschke, R.~Garnavi,
  Semi-supervised segmentation of optic cup in retinal fundus images using
  variational autoencoder,, in: In Proc. MICCAI, 2017, pp. 75--82 (2017).

\bibitem{FlowNet}
A.~Dosovitskiy, P.~Fischer, et. al., Flownet: Learning optical flow with
  convolutional networks, in: In Proc. IEEE ICCV, 2015, pp. 2758--2766 (2015).

\bibitem{Mahapatra_MICCAI17}
D.~Mahapatra, S.~Bozorgtabar, S.~Hewavitahranage, R.~Garnavi, Image super
  resolution using generative adversarial networks and local saliencymaps for
  retinal image analysis,, in: In Proc. MICCAI, 2017, pp. 382--390 (2017).

\bibitem{Roy_ISBI17}
P.~Roy, R.~Tennakoon, K.~Cao, S.~Sedai, D.~Mahapatra, S.~Maetschke, R.~Garnavi,
  A novel hybrid approach for severity assessment of diabetic retinopathy in
  colour fundus images,, in: In Proc. IEEE ISBI, 2017, pp. 1078--1082 (2017).

\bibitem{Roy_DICTA16}
P.~Roy, R.~Chakravorty, S.~Sedai, D.~Mahapatra, R.~Garnavi, Automatic eye type
  detection in retinal fundus image using fusion of transfer learning and
  anatomical features, in: In Proc. DICTA, 2016, pp. 1--7 (2016).

\bibitem{Tennakoon_OMIA16}
R.~Tennakoon, D.~Mahapatra, P.~Roy, S.~Sedai, R.~Garnavi, Image quality
  classification for dr screening using convolutional neural networks, in: In
  Proc. MICCAI-OMIA, 2016, pp. 113--120 (2016).

\bibitem{Sedai_OMIA16}
S.~Sedai, P.~Roy, D.~Mahapatra, R.~Garnavi, Segmentation of optic disc and
  optic cup in retinal images using coupled shape regression, in: In Proc.
  MICCAI-OMIA, 2016, pp. 1--8 (2016).

\bibitem{STN}
M.~Jaderberg, K.~Simonyan, A.~Zisserman, K.~Kavukcuoglu, Spatial transformer
  networks, in: NIPS, 2015, pp.~-- (2015).

\bibitem{Mahapatra_MLMI16}
D.~Mahapatra, P.~Roy, S.~Sedai, R.~Garnavi, Retinal image quality
  classification using saliency maps and cnns, in: In Proc. MICCAI-MLMI, 2016,
  pp. 172--179 (2016).

\bibitem{Sedai_EMBC16}
S.~Sedai, P.~Roy, D.~Mahapatra, R.~Garnavi, Segmentation of optic disc and
  optic cup in retinal fundus images using shape regression, in: In Proc. EMBC,
  2016, pp. 3260--3264 (2016).

\bibitem{Mahapatra_EMBC16}
D.~Mahapatra, P.~Roy, S.~Sedai, R.~Garnavi, A cnn based neurobiology inspired
  approach for retinal image quality assessment, in: In Proc. EMBC, 2016, pp.
  1304--1307 (2016).

\bibitem{Mahapatra_MLMI15_Optic}
J.~Zilly, J.~Buhmann, D.~Mahapatra, Boosting convolutional filters with entropy
  sampling for optic cup and disc image segmentation from fundus images, in: In
  Proc. MLMI, 2015, pp. 136--143 (2015).

\bibitem{Mahapatra_MLMI15_Prostate}
D.~Mahapatra, J.~Buhmann, Visual saliency based active learning for prostate
  mri segmentation, in: In Proc. MLMI, 2015, pp. 9--16 (2015).

\bibitem{Mahapatra_OMIA15}
D.~Mahapatra, J.~Buhmann, Obtaining consensus annotations for retinal image
  segmentation using random forest and graph cuts, in: In Proc. OMIA, 2015, pp.
  41--48 (2015).

\bibitem{MahapatraISBI15_Optic}
D.~Mahapatra, J.~Buhmann, A field of experts model for optic cup and disc
  segmentation from retinal fundus images, in: In Proc. IEEE ISBI, 2015, pp.
  218--221 (2015).

\bibitem{MahapatraISBI15_JSGR}
D.~Mahapatra, Z.~Li, F.~Vos, J.~Buhmann, Joint segmentation and groupwise
  registration of cardiac dce mri using sparse data representations, in: In
  Proc. IEEE ISBI, 2015, pp. 1312--1315 (2015).

\bibitem{MahapatraISBI15_CD}
D.~Mahapatra, F.~Vos, J.~Buhmann, Crohn's disease segmentation from mri using
  learned image priors, in: In Proc. IEEE ISBI, 2015, pp. 625--628 (2015).

\bibitem{KuangAMM14}
H.~Kuang, B.~Guthier, M.~Saini, D.~Mahapatra, A.~E. Saddik, A real-time smart
  assistant for video surveillance through handheld devices., in: In Proc: ACM
  Intl. Conf. Multimedia, 2014, pp. 917--920 (2014).

\bibitem{Vos_DIR}
B.~de~Vos, F.~Berendsen, M.~Viergever, M.~Staring, I.~Isgum, End-to-end
  unsupervised deformable image registration with a convolutional neural
  network, in: MICCAI International Workshop on Deep Learning in Medical Image
  Analysis, 2017, pp. 204--212 (2017).

\bibitem{Mahapatra_ABD2014}
D.~Mahapatra, J.Tielbeek, J.~Makanyanga, J.~Stoker, S.~Taylor, F.~Vos,
  J.~Buhmann, Combining multiple expert annotations using semi-supervised
  learning and graph cuts for crohn's disease segmentation, in: In Proc:
  MICCAI-ABD, 2014 (2014).

\bibitem{Schuffler_ABD2014}
P.~Sch$\ddot{u}$ffler, D.~Mahapatra, J.~Tielbeek, F.~Vos, J.~Makanyanga,
  D.~Pends, C.~Nio, J.~Stoker, S.~Taylor, J.~Buhmann, Semi automatic crohns
  disease severity assessment on mr imaging, in: In Proc: MICCAI-ABD, 2014
  (2014).

\bibitem{MahapatraISBI_CD2014}
D.~Mahapatra, J.Tielbeek, J.~Makanyanga, J.~Stoker, S.~Taylor, F.~Vos,
  J.~Buhmann, Active learning based segmentation of crohn's disease using
  principles of visual saliency, in: Proc. IEEE ISBI, 2014, pp. 226--229
  (2014).

\bibitem{MahapatraMICCAI_CD2013}
D.~Mahapatra, P.~Sch$\ddot{u}$ffler, J.~Tielbeek, F.~Vos, J.~Buhmann,
  Semi-supervised and active learning for automatic segmentation of crohn's
  disease, in: Proc. MICCAI, Part 2, 2013, pp. 214--221 (2013).

\bibitem{Schuffler_ABD2013}
P.~Sch$\ddot{u}$ffler, D.~Mahapatra, J.~Tielbeek, F.~Vos, J.~Makanyanga,
  D.~Pends, C.~Nio, J.~Stoker, S.~Taylor, J.~Buhmann, A model development
  pipeline for crohns disease severity assessment from magnetic resonance
  images, in: In Proc: MICCAI-ABD, 2013 (2013).

\bibitem{UzunovaMICCAI2017}
H.~Uzunova, M.~Wilms, H.~Handels, J.~Ehrhardt, Training cnns for image
  registration from few samples with model-based data augmentation, in: In
  Proc. MICCAI, 2017, pp. 223--231 (2017).

\bibitem{MahapatraProISBI13}
D.~Mahapatra, Graph cut based automatic prostate segmentation using learned
  semantic information, in: Proc. IEEE ISBI, 2013, pp. 1304--1307 (2013).

\bibitem{MahapatraRVISBI13}
D.~Mahapatra, J.~Buhmann, Automatic cardiac rv segmentation using semantic
  information with graph cuts, in: Proc. IEEE ISBI, 2013, pp. 1094--1097
  (2013).

\bibitem{MahapatraWssISBI13}
D.~Mahapatra, J.~Tielbeek, F.~Vos, J.~Buhmann, Weakly supervised semantic
  segmentation of crohn's disease tissues from abdominal mri, in: Proc. IEEE
  ISBI, 2013, pp. 832--835 (2013).

\bibitem{MahapatraCDFssISBI13}
D.~Mahapatra, J.~Tielbeek, F.~Vos, J.~B. ., Crohn's disease tissue segmentation
  from abdominal mri using semantic information and graph cuts, in: Proc. IEEE
  ISBI, 2013, pp. 358--361 (2013).

\bibitem{MahapatraCDSPIE13}
D.~Mahapatra, J.~Tielbeek, F.~Vos, J.~Buhmann, Localizing and segmenting
  crohn's disease affected regions in abdominal mri using novel context
  features, in: Proc. SPIE Medical Imaging, 2013 (2013).

\bibitem{RoheMICCAI2017}
M.~Rohe, M.~Datar, T.~Heimann, M.~Sermesant, X.~Pennec, {SVF-N}et: Learning
  deformable image registration using shape matching, in: In Proc.MICCAI, 2017,
  pp. 266--274 (2017).

\bibitem{MahapatraABD12}
D.~Mahapatra, J.~Tielbeek, J.~Buhmann, F.~Vos, A supervised learning based
  approach to detect crohn's disease in abdominal mr volumes, in: Proc. MICCAI
  workshop Computational and Clinical Applications in Abdominal
  Imaging(MICCAI-ABD), 2012, pp. 97--106 (2012).

\bibitem{MahapatraMLMI12}
D.~Mahapatra, Cardiac lv and rv segmentation using mutual context information,
  in: Proc. MICCAI-MLMI, 2012, pp. 201--209 (2012).

\bibitem{MahapatraSTACOM12}
D.~Mahapatra, Landmark detection in cardiac mri using learned local image
  statistics, in: Proc. MICCAI-Statistical Atlases and Computational Models of
  the Heart. Imaging and Modelling Challenges (STACOM), 2012, pp. 115--124
  (2012).

\bibitem{VosEMBC}
F.~M. Vos, J.~Tielbeek, R.~Naziroglu, Z.~Li, P.~Sch$\ddot{u}$ffler,
  D.~Mahapatra, A.~Wiebel, C.~Lavini, J.~Buhmann, H.~Hege, J.~Stoker, L.~van
  Vliet, Computational modeling for assessment of {IBD}: to be or not to be?,
  in: Proc. IEEE EMBC, 2012, pp. 3974--3977 (2012).

\bibitem{MahapatraGRSPIE12}
D.~Mahapatra, Groupwise registration of dynamic cardiac perfusion images using
  temporal information and segmentation information, in: In Proc: SPIE Medical
  Imaging, 2012 (2012).

\bibitem{MahapatraMiccaiIAHBD11}
D.~Mahapatra, Neonatal brain mri skull stripping using graph cuts and shape
  priors, in: In Proc: MICCAI workshop on Image Analysis of Human Brain
  Development (IAHBD), 2011 (2011).

\bibitem{MahapatraMiccai11}
D.~Mahapatra, Y.~Sun, Orientation histograms as shape priors for left ventricle
  segmentation using graph cuts, in: In Proc: MICCAI, 2011, pp. 420--427
  (2011).

\bibitem{MahapatraMiccai10}
D.~Mahapatra, Y.~Sun, Joint registration and segmentation of dynamic cardiac
  perfusion images using mrfs., in: Proc. MICCAI, 2010, pp. 493--501 (2010).

\bibitem{MahapatraICIP10}
D.~Mahapatra, Y.~Sun., An mrf framework for joint registration and segmentation
  of natural and perfusion images, in: Proc. IEEE ICIP, 2010, pp. 1709--1712
  (2010).

\bibitem{MahapatraICDIP10a}
D.~Mahapatra, Y.~Sun, Retrieval of perfusion images using cosegmentation and
  shape context information, in: Proc. APSIPA Annual Summit and Conference
  (ASC), 2010 (2010).

\bibitem{MahapatraICDIP10b}
D.~Mahapatra, Y.~Sun, A saliency based mrf method for the joint registration
  and segmentation of dynamic renal mr images, in: Proc. ICDIP, 2010 (2010).

\bibitem{MahapatraMiccai08}
D.~Mahapatra, Y.~Sun, Nonrigid registration of dynamic renal {MR} images using
  a saliency based {MRF} model, in: Proc. MICCAI, 2008, pp. 771--779 (2008).

\bibitem{MahapatraISBI08}
D.~Mahapatra, Y.~Sun, Registration of dynamic renal {MR} images using
  neurobiological model of saliency, in: Proc. ISBI, 2008, pp. 1119--1122
  (2008).

\bibitem{MahapatraICME08}
D.~Mahapatra, M.~Saini, Y.~Sun, Illumination invariant tracking in office
  environments using neurobiology-saliency based particle filter, in: IEEE
  ICME, 2008, pp. 953--956 (2008).

\bibitem{MahapatraICBME08_Retrieve}
D.~Mahapatra, S.~Roy, Y.~Sun, Retrieval of mr kidney images by incorporating
  spatial information in histogram of low level features, in: In 13th
  International Conference on Biomedical Engineering, 2008 (2008).

\bibitem{MahapatraICBME08_Sal}
D.~Mahapatra, Y.~Sun, Using saliency features for graphcut segmentation of
  perfusion kidney images, in: In 13th International Conference on Biomedical
  Engineering, 2008 (2008).

\bibitem{MahapatraSPIE08}
D.~Mahapatra, S.~Winkler, S.~Yen, Motion saliency outweighs other low-level
  features while watching videos, in: SPIE HVEI., 2008, pp. 1--10 (2008).

\bibitem{MahapatraICIT06}
D.~Mahapatra, A.~Routray, C.~Mishra, An active snake model for classification
  of extreme emotions, in: IEEE International Conference on Industrial
  Technology (ICIT), 2006, pp. 2195--2199 (2006).

\bibitem{MahapatraGANISBI2018}
D.~Mahapatra, B.~Antony, S.~Sedai, R.~Garnavi, Deformable medical image
  registration using generative adversarial networks, in: In Proc. IEEE ISBI,
  2018, pp. 1449--1453 (2018).

\bibitem{sZoom_Ar}
M.~Saini, B.~Guthier, H.~Kuang, D.~Mahapatra, A.~Saddik, szoom: A framework for
  automatic zoom into high resolution surveillance videos, in: arXiv preprint
  arXiv:1909.10164, 2019 (2019).

\bibitem{CVIU_Ar}
B.~Bozorgtabar, D.~Mahapatra, H.~von Teng, A.~Pollinger, L.~Ebner, J.-P.
  Thiran, M.~Reyes, Informative sample generation using class aware generative
  adversarial networks for classification of chest xrays, in: arXiv preprint
  arXiv:1904.10781, 2019 (2019).

\bibitem{AMD_OCT}
D.~Mahapatra, Amd severity prediction and explainability using image
  registration and deep embedded clustering, in: arXiv preprint
  arXiv:1907.03075, 2019 (2019).

\bibitem{GANReg1_Ar}
D.~Mahapatra, Z.~Ge, Combining transfer learning and segmentation information
  with gans for training data independent image registration, in: arXiv
  preprint arXiv:1903.10139, 2019 (2019).

\bibitem{PGAN_Ar}
D.~Mahapatra, B.~Bozorgtabar, Progressive generative adversarial networks for
  medical image super resolution, in: arXiv preprint arXiv:1902.02144, 2019
  (2019).

\bibitem{Haze_Ar}
S.~Kuanar, K.~Rao, D.~Mahapatra, M.~Bilas, Night time haze and glow removal
  using deep dilated convolutional network, in: arXiv preprint
  arXiv:1902.00855, 2019 (2019).

\bibitem{Xr_Ar}
Z.~Ge, D.~Mahapatra, S.~Sedai, R.~Garnavi, R.~Chakravorty, Chest x-rays
  classification: A multi-label and fine-grained problem, in: arXiv preprint
  arXiv:1807.07247, 2018 (2018).

\bibitem{RegGan_Ar}
D.~Mahapatra, S.~Sedai, R.~Garnavi, Elastic registration of medical images with
  gans, in: arXiv preprint arXiv:1805.02369, 2018 (2018).

\bibitem{ISR_Ar}
D.~Mahapatra, B.~Bozorgtabar, Retinal vasculature segmentation using local
  saliency maps and generative adversarial networks for image super resolution,
  in: arXiv preprint arXiv:1710.04783, 2017 (2017).

\bibitem{LME_Ar}
D.~Mahapatra, Consensus based medical image segmentation using semi-supervised
  learning and graph cuts, in: arXiv preprint arXiv:1612.02166, 2017 (2017).

\bibitem{Misc}
D.~Mahapatra, K.~Agarwal, R.~Khosrowabadi, D.~Prasad, Recent advances in
  statistical data and signal analysis: Application to real world diagnostics
  from medical and biological signals, in: Computational and mathematical
  methods in medicine, 2016 (2016).

\bibitem{Health_p}
P.~Bastide, I.~Kiral-Kornek, D.~Mahapatra, S.~Saha, A.~Vishwanath, S.~V.
  Cavallar, Crowdsourcing health improvements routes, in: US Patent App.
  15/611,519, 2019 (2019).

\bibitem{Pat2}
D.~Mahapatra, R.~Garnavi, P.~Roy, R.~Tennakoon, System and method to teach and
  evaluate image grading performance using prior learned expert knowledge base,
  in: US Patent App. 15/459,457, 2018 (2018).

\bibitem{Pat3}
D.~Mahapatra, R.~Garnavi, P.~Roy, R.~Tennakoon, System and method to teach and
  evaluate image grading performance using prior learned expert knowledge base,
  in: US Patent App. 15/814,590, 2018 (2018).

\bibitem{Pat4}
D.~Mahapatra, R.~Garnavi, S.~Sedai, R.~Tennakoon, Generating an enriched
  knowledge base from annotated images, in: US Patent App. 15/429,735, 2018
  (2018).

\bibitem{PR_DLFace}
C.~Peng, N.~Wang, J.~Li, X.~Gao, Dlface: Deep local descriptor for
  cross-modality face recognition., Pattern Recognition 90 (2019) 161–171
  (2019).

\bibitem{Pat5}
P.~Bastide, I.~Kiral-Kornek, D.~Mahapatra, S.~Saha, A.~Vishwanath, S.~V.
  Cavallar, Visual health maintenance and improvement, in: US Patent 9,993,385,
  2018 (2018).

\bibitem{Pat6}
D.~Mahapatra, R.~Garnavi, S.~Sedai, R.~Tennakoon, Classification of severity of
  pathological condition using hybrid image representation, in: US Patent App.
  15/426,634, 2018 (2018).

\bibitem{Pat7}
P.~Bastide, I.~Kiral-Kornek, D.~Mahapatra, S.~Saha, A.~Vishwanath, S.~V.
  Cavallar, Machine learned optimizing of health activity for participants
  during meeting times, in: US Patent App. 15/426,634, 2018 (2018).

\bibitem{Pat8}
D.~Mahapatra, R.~Garnavi, S.~Sedai, R.~Tennakoon, R.~Chakravorty, Early
  prediction of age related macular degeneration by image reconstruction, in:
  US Patent App. 15/854,984, 2018 (2018).

\bibitem{Pat9}
D.~Mahapatra, R.~Garnavi, S.~Sedai, R.~Tennakoon, R.~Chakravorty, Early
  prediction of age related macular degeneration by image reconstruction, in:
  US Patent 9,943,225, 2018 (2018).

\bibitem{PR_Shang}
R.~Shang, Y.~Meng, W.~Wang, F.~Shang, L.~Jiao, Local discriminative based
  sparse subspace learning for feature selection., Pattern Recognition 92
  (2019) 219--230 (2019).

\bibitem{Pat10}
D.~Mahapatra, R.~Garnavi, S.~Sedai, P.~Roy, Retinal image quality assessment,
  error identification and automatic quality correction, in: US Patent
  9,779,492, 2017 (2017).

\bibitem{Pat11}
D.~Mahapatra, R.~Garnavi, S.~Sedai, P.~Roy, Joint segmentation and
  characteristics estimation in medical images, in: US Patent App. 15/234,426,
  2017 (2017).

\bibitem{PR_Yang}
X.~Yang, W.~Wu, Y.~Chen, X.~Li, J.~Zhang, D.~Long, L.~Yang, An integrated
  inverse space sparse representation framework for tumor classification.,
  Pattern Recognition 93 (2019) 293--311 (2019).

\bibitem{PR_Lei}
B.~Lei, K.~Jinman, E.~Ahn, A.~Kumar, F.~Dagan, M.~Fulham, Step-wise integration
  of deep class-specific learning for dermoscopic image segmentation., Pattern
  Recognition 85 (2019) 78--89 (2019).

\bibitem{PR_HYang}
H.~Yang, C.~Yuan, B.~Li, Y.~Du, J.~Xing, W.~Hu, S.~Maybank, Asymmetric 3d
  convolutional neural networks for action recognition., Pattern Recognition 85
  (2019) 1--12 (2019).

\bibitem{PR_WWang}
W.~Wang, H.~Wang, Z.~Zhang, C.~Zhang, Y.~Gao, Semi-supervised domain adaptation
  via fredholm integral based kernel methods., Pattern Recognition 85 (2019)
  185--197 (2019).

\bibitem{PR_LHou}
L.~Hou, V.~Nguyen, A.B.Kanevsky, D.~Samaras, T.~M. Kurc, T.~Zhao, R.~Gupta,
  Y.~Gao, W.~Chen, D.~Foran, J.~Saltz, Sparse autoencoder for unsupervised
  nucleus detection and representation in histopathology images., Pattern
  Recognition 86 (2019) 188--200 (2019).

\bibitem{GANs}
I.~Goodfellow, J.~Pouget-Abadie, M.~Mirza, B.~Xu, D.~Warde-Farley, S.~Ozair,
  A.~Courville, Y.~Bengio, Generative adversarial nets, in: Proc. NIPS, 2014,
  pp. 2672--2680 (2014).

\bibitem{SRGAN}
C.~Ledig, L.~Theis, F.~Huszár, J.~Caballero, A.~Cunningham, A.~Acosta,
  A.~Aitken, A.~Tejani, J.~Totz, Z.~Wang, W.~Shi, Photo-realistic single image
  super-resolution using a generative adversarial network, in: CVPR, 2017, pp.
  4681--4690 (2017).

\bibitem{MahapatraMICCAI_ISR}
D.~Mahapatra, B.~Bozorgtabar, S.~Hewavitharanage, R.~Garnavi, Image super
  resolution using generative adversarial networks and local saliency maps for
  retinal image analysis, in: MICCAI, 2017, pp. 382--390 (2017).

\bibitem{CondGANs}
P.~Isola, J.~Zhu, T.~Zhou, A.~Efros, Image-to-image translation with
  conditional adversarial networks, in: IEEE CVPR, 2017 (2017).

\bibitem{CyclicGANs}
J.~Zhu, T.~Park, P.~Isola, A.~Efros, Unpaired image-to-image translation using
  cycle-consistent adversarial networks, in: IEEE ICCV, 2017, pp. 2223--2232
  (2017).

\bibitem{FFD}
D.~Rueckert, L.~Sonoda, C.~Hayes, D.~Hill, M.~Leach, D.~Hawkes., Nonrigid
  registration using free-form deformations: application to breast mr images.,
  IEEE Trans. Med. Imag.. 18~(8) (1999) 712--721 (1999).

\bibitem{SSIM}
Z.~Wang, et. al., Image quality assessment: from error visibility to structural
  similarity., IEEE Trans. Imag. Proc. 13~(4) (2004) 600--612 (2004).

\bibitem{VGG}
K.~Simonyan, A.~Zisserman., Very deep convolutional networks for large-scale
  image recognition, CoRR abs/1409.1556 (2014).

\bibitem{NIHXray}
X.~Wang, Y.~Peng, L.~Lu, Z.~Lu, M.~Bagheri, R.~Summers, Chestx-ray8:
  Hospital-scale chest x-ray database and benchmarks on weakly-supervised
  classification and localization of common thorax diseases, in: In Proc. CVPR,
  2017 (2017).

\bibitem{Adam}
D.~Kingma, J.~Ba, Adam: A method for stochastic optimization, in: International
  Conference on Learning Representations,, 2014 (2014).

\bibitem{Elastix}
S.~Klein, M.~Staring, K.~Murphy, M.~Viergever, J.~Pluim., Elastix: a toolbox
  for intensity based medical image registration., IEEE Trans. Med. Imag..
  29~(1) (2010) 196--205 (2010).

\bibitem{Imagenet}
J.~Deng, W.~Dong, R.~Socher, L.-J. Li, K.~Li, L.~Fei-Fei., {ImageNet: A
  Large-Scale Hierarchical Image Database}, in: CVPR09, 2009 (2009).

\bibitem{Bala33}
S.~G. Mueller, Ways toward an early diagnosis in alzheimer’s disease: the
  alzheimer’s disease neuroimaging initiative ({ADNI})., Alzheimer’s \&
  Dementia 1~(1) (2005) 55--66 (2005).

\bibitem{Bala17}
B.~Fischl, Freesurfer, Neuroimage 62~(2) (2015) 774--781 (2015).

\bibitem{Alipour2014}
S.~Hajeb, H.~Rabbani, M.~Akhlaghi., Diabetic retinopathy grading by digital
  curvelet transform., Comput Math Methods Med. (2012) 7619--01 (2012).

\bibitem{Frangi1998}
A.~Frangi, W.~Niessen, K.~Vincken, M.~Viergever, Multiscale vessel enhancement
  filtering, in: MICCAI, 1998, pp. 130--137 (1998).

\end{thebibliography}

\end{document}